\documentclass{cDSS2e}  
  
\usepackage{amsfonts}  
\usepackage{amsmath}  
\usepackage{xspace}  
\usepackage{epsf}  
  
\newcommand{\Rset}{\mathbb{R}}  
\newcommand{\sgn}{\mathop{\mathrm{sgn}}}  
  
\newcommand{\smallin}{{\mbox{\footnotesize in}}}  
\newcommand{\smallout}{{\mbox{\footnotesize out}}}  
\newcommand{\smalli}{{\mbox{\footnotesize i}}}  
\newcommand{\smallr}{{\mbox{\footnotesize r}}}  
\newcommand{\smallu}{{\mbox{\footnotesize u}}}  
\newcommand{\smalls}{{\mbox{\footnotesize s}}}

\begin{document}  
\doi{10.1080/14689360xxxxxxxxxxxxx}  
 \issn{1468-9375}  
\issnp{1468-9367} \jvol{00} \jnum{00} \jyear{2007} \jmonth{June}  
\markboth{Kirk and Rucklidge}{Effect of symmetry breaking}

\title{The effect of symmetry breaking on the dynamics near a structurally  
stable heteroclinic cycle between equilibria and a periodic orbit}  
  
\author{Vivien Kirk,\thanks{V.Kirk@auckland.ac.nz} 
Department of Mathematics, University of Auckland,  
\break Private Bag 92019, Auckland, New Zealand  
\break Alastair M. Rucklidge,\thanks{A.M.Rucklidge@leeds.ac.uk} 
Department of Applied Mathematics, University of Leeds,  
Leeds LS2 9JT, UK}  
  
\received{\today}  
\maketitle  
  
\begin{abstract}  
The effect of small forced symmetry breaking on the dynamics near a 
structurally stable heteroclinic cycle connecting two equilibria and a periodic 
orbit is investigated.  This type of system is known to exhibit complicated, 
possibly chaotic dynamics including irregular switching of sign of various 
phase space variables, but details of the mechanisms underlying the complicated 
dynamics have not previously been investigated.  We identify global 
bifurcations that induce the onset of chaotic dynamics and switching near a 
heteroclinic cycle of this type, and by construction and analysis of 
approximate return maps, locate the global bifurcations in parameter space. We 
find there is a threshold in the size of certain symmetry-breaking terms below 
which there can be no persistent switching. Our results are illustrated by a 
numerical example. 
 \end{abstract}  
  
\section{Introduction}  
  
It is well-established that the presence of symmetries in dynamical systems can
result in the existence of heteroclinic cycles that are structurally stable
with respect to symmetric perturbations~\cite{F80,GuHo88}. By {\em heteroclinic
cycle} we mean a collection of two or more flow invariant sets $\{\xi_1,\dots
,\xi_n\}$ of some system of ordinary differential equations together with a set
of heteroclinic connections $\{\gamma_1(t), \dots,\gamma_n(t)\}$, where
$\gamma_j(t) \rightarrow \xi_{j}$ as $t \rightarrow -\infty$ and $\gamma_j(t)
\rightarrow \xi_{j+1}$ as $t \rightarrow +\infty$, and where $\xi_{n+1}\equiv
\xi_1$. In many studies, all the $\xi_i$ are equilibria, but in this paper we
explicitly consider the case that one of the $\xi_i$ is a periodic orbit. The
connections $\gamma_i$ may be isolated, or there may be a continuum of
connections from $\xi_i$ to $\xi_{i+1}$ for one or more $i$.
  
There is a large literature on structurally stable heteroclinic cycles (SSHC),  
including work establishing conditions for the existence and asymptotic  
stability of heteroclinic cycles~\cite{KrMe95,KrMe04,M91}, examination of the  
dynamics near heteroclinic cycles and networks of heteroclinic cycles  
\cite{KS94,AC98,ACL05,PD05}, and unfolding of bifurcations of heteroclinic  
cycles~\cite{SC92,CKMS,PD05b}. SSHC arise naturally in mathematical models of  
physical systems  with symmetry or near-symmetry~\cite{BuHe80,AGH88,PJ88,NTDX}. In  
these models, the physical system is idealised as having perfect symmetry,  
leading to the existence of invariant subspaces in the model and thus to the  
robustness of heteroclinic cycles with respect to symmetric perturbations. It  
is natural to ask how much of the dynamics observed in symmetric models  
persists under non-symmetric perturbations. Some effects of small  
symmetry-breaking have been documented~\cite{Me89,Ch93,SaSc95,MPR}, and aspects  
of the related question of how much of the dynamics persists under the  
inclusion of small noise have also been considered~\cite{StHo90,ASK03}, but  
details are likely to vary greatly between different examples. A few cases of 
experimental observation of near-heteroclinic cycles have been reported, 
most recently in  
\cite{NMQ}, but see also the references therein. In these cases, experimental noise and small 
symmetry-breaking effects prevent exact heteroclinic cycles from occurring, 
but there is clear evidence for near-heteroclinic structures in certain regimes. 
  
\begin{figure}  
 \centerline{\epsfxsize=0.9\hsize\epsfbox{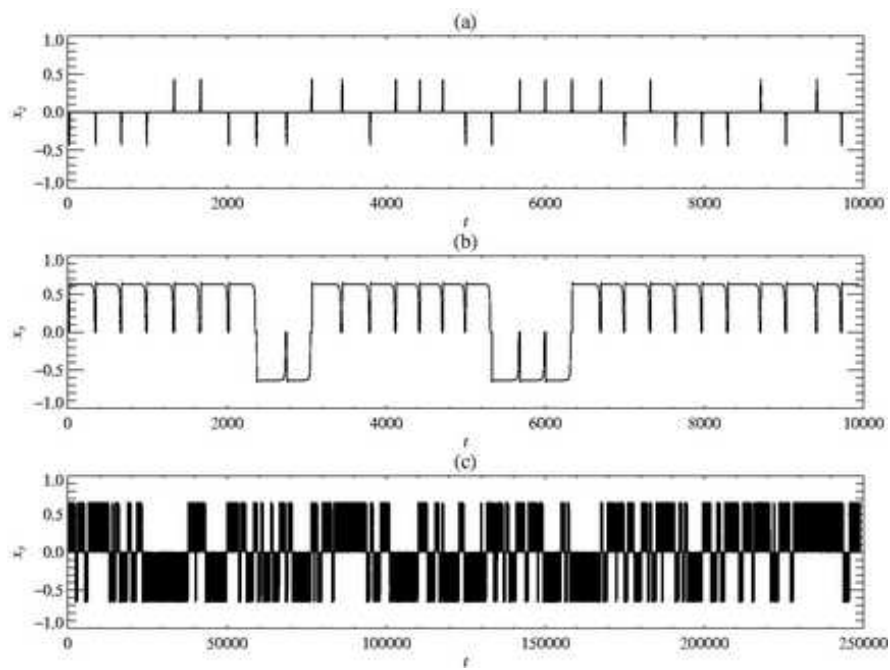}}  
 \caption{Irregular switching in the time series of a dynamo model studied 
in~\cite[figure~1]{MPR}. Panels~(a) and (b) show the evolution of different 
coordinates of the same trajectory, and panel~(c) shows the same coordinate as 
in~(b) over a longer time interval.} 
 \label{fig:MPRexample}  
 \end{figure}

Our interest in the particular set-up explored in this paper is motivated by  
\cite{MPR}, which makes the observation that the addition of small  
symmetry-breaking terms to a system containing a heteroclinic cycle connecting  
two equilibria and a periodic orbit (as well as symmetric copies of the cycle)  
results in seemingly chaotic dynamics, with orbits passing near the various  
equilibria in the system repeatedly but in an irregular pattern, as illustrated  
in figure~\ref{fig:MPRexample}. A main point of~\cite{MPR} was to show that  
repeated switching of orbits in this manner could arise in a simple  
four-dimensional, nearly symmetric model, but the specific mechanisms  
underlying the complicated dynamics were not explored in detail.  
  
 %%%%%%%%%%%%%%%%  
 %%%%%%%%%%%%%%%%%  
  
In this paper, we examine a generalisation of the situation from~\cite{MPR}, 
focusing on the structure and origin of chaotic dynamics in the system and on 
how switching dynamics is induced. Here and elsewhere in the paper, {\em 
switching} refers to the itinerary that an orbit follows under the dynamics. 
Specifically, in the fully symmetric version of our system there is a 
heteroclinic network consisting of four symmetric copies of the basic 
heteroclinic cycle. Invariance of various subspaces ensures that an orbit may 
make repeated passes near only one cycle. Once the symmetries are broken, 
however, an orbit may {\em switch}, i.e., make traversals near more than one of 
the original cycles (although, of course, the cycles themselves may not persist 
when the symmetry is broken). 
  
A main result of this paper is that in the case of small symmetry breaking,  
switching in one variable occurs when a complicated attractor arising from the  
presence of transverse homoclinic orbits of a periodic orbit crosses the stable  
manifold of one of the equilibria in the system. The existence of the  
transverse homoclinic orbits depends on a broken rotation 
symmetry, while the proximity of the attractor to the stable manifold of the  
equilibrium is caused by a broken reflection symmetry. Switching in a second  
variable results from the interaction between broken reflection symmetry and  
complicated dynamics associated with a heteroclinic bifurcation between the  
equilibria. Thus, switching results from the right combination of a global  
bifurcation and small symmetry breaking.  
  
A second significant result of this paper is the observation that there is a 
threshold for the size of symmetry breaking below which 
persistent switching cannot occur. More precisely, the existence of the  
heteroclinic cycle requires three separate symmetries to allow structurally  
stable connections within three invariant subspaces. We control the degree to  
which the three symmetries are broken by three small parameters, $\epsilon_1$,  
$\epsilon_2$ and~$\epsilon_3$; $\epsilon_1$~controls the degree  
to which the periodic orbit in the cycle deviates from a perfect circle, while  
$\epsilon_2$ and $\epsilon_3$ break reflection symmetries. For fixed small  
$\epsilon_2$ and~$\epsilon_3$, we find that there is a threshold  
in~$\epsilon_1$ for persistent switching to occur. For sufficiently  
small~$\epsilon_1$, there may be a single switch from one part of phase space  
to another, but it is only for $\epsilon_1$ beyond the threshold value that an 
orbit can repeatedly visit different parts of the phase space. We find that  
it is possible to get sustained switching in one or other or both of the  
variables associated with the reflection symmetries, and that the threshold  
values of~$\epsilon_1$ are different for switching in the two variables. The threshold does  
not go to zero as $\epsilon_2$ and~$\epsilon_3$ go to zero.  
  
Sustained switching of orbits near heteroclinic cycles and networks has been  
observed in a number of other settings. Clune and Knobloch~\cite{ClKn94}  
describe an example in which there are two symmetrically related copies of a  
non-asymptotically stable heteroclinic cycle, with nearby orbits making  
repeated passes near each cycle; no mechanism for the switching is suggested in  
this paper.  Aguiar et al.~\cite{ACL05} find switching near a hybrid  
heteroclinic network formed from transverse heteroclinic connections between  
equilibria and connections that are robust because of symmetry; switching seems  
to result from the folding and stretching caused by passage near the  
transversal heteroclinic connections and by mixing near an equilibrium solution  
with complex eigenvalues. Kirk et al.~\cite{KLS05} have an example of switching  
near a heteroclinic network that has no transversal connections; the switching  
is caused entirely by passage near an equilibrium with complex eigenvalues.  
Postlethwaite and Dawes~\cite{PD05} describe a variant of switching near a  
heteroclinic network in which each cycle in the network is unstable along a  
direction transverse to the cycle; orbits visit cycles in the network in a  
fixed order (being pushed away from each cycle in the transverse direction,  
which also happens to be the contracting direction for the next cycle) but the  
number of traversals of each cycle before switching to the next cycle can be  
constant or irregular. Ashwin et al.~\cite{ARS04a} describe switching  
associated with a stuck-on heteroclinic cycle between two invariant subspaces;  
here the switching is caused by a nonlinear mechanism that chooses between the different  
possibilities in a manner that is well modelled by a random process. Switching  
can also be induced by adding noise to a structurally stable heteroclinic  
network~\cite{ASK03}; noise sensitive switching has been observed  
by~\cite{RM95,MRWP96}. None of these examples explicitly considers symmetry  
breaking as a mechanism for switching.  
  
We adopt a standard approach to analysis of the system of interest, i.e., we  
set up a simple symmetric model in which there exists a heteroclinic cycle  
connecting two equilibria and a periodic orbit (Section~\ref{sec:description}),  
construct a return map that approximates the dynamics near such a cycle, and  
then add generic symmetry breaking terms to the return map  
(Section~\ref{sec:construct}, with details in the Appendix).  
Analysis of the return map is fruitful in cases  
where partial symmetry is retained, and allows us to prove the existence and  
asymptotic stability of periodic orbits, quasiperiodic solutions or  
heteroclinic cycles in various cases  
(Sections~\ref{sec:global}--\ref{sec:rotate}). In the completely asymmetric  
case, the return map is intractable, but we are able to make predictions about  
the dynamics by assuming there is a generic unfolding of the partially  
symmetric cases (Section~\ref{sec:full}). The example discussed in  
Section~\ref{sec:numerics} confirms and illustrates the analysis. Some  
conclusions are presented in Section~\ref{sec:conclusions}.  
  
A complicating factor in the analysis presented in this paper is that the  
unstable manifolds of one pair of equilibria and of the periodic orbit are  
two-dimensional, and there are continua of heteroclinic connections along some  
parts of the cycle in the fully symmetric case. Linearising about a single  
heteroclinic connection is not appropriate, and the usual method of analysis  
needs to be adapted to keep track of orbits in a neighbourhood of all the  
connections. Our approach is similar to that taken in~\cite{AC98,R01,KLS05}. We  
note that our analysis need not consider the issue of which connection from a  
continuum is selected by the dynamics (as investigated in, for instance,  
\cite{AC98,AFRS03,ARS04a,ARS04b}) since in our case breaking of the symmetries  
forces a discrete set of transversal connections to be selected from each  
continuum. Note also that some results about the dynamics near a heteroclinic 
cycle connecting an equilibrium and a periodic orbit in a generic (i.e., non-symmetric) 
setting are described in \cite{Rademacher, KROCK}, but the  
phenomena described  
in those papers  
will not be seen for small symmetry breaking in our setting, 
and is not the focus of our interest here. 
  
\section{Description of the problem}  
\label{sec:description}  
  
\begin{figure}  
 \centerline{\makebox[0.3\hsize][l]{(a)}  
             \hspace{0.03\hsize}  
             \makebox[0.3\hsize][l]{(b)} \hspace{0.03\hsize}  
             \makebox[0.3\hsize][l]{(c)}}  
 \vspace{0.3ex}  
 \centerline{\epsfxsize=0.3\hsize\epsfbox{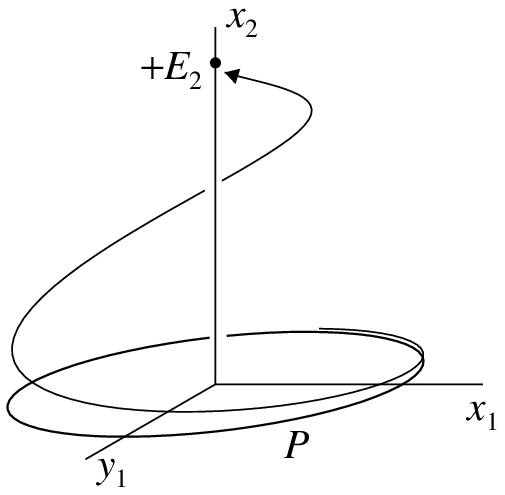}  
             \hspace{0.03\hsize}  
             \epsfxsize=0.3\hsize\epsfbox{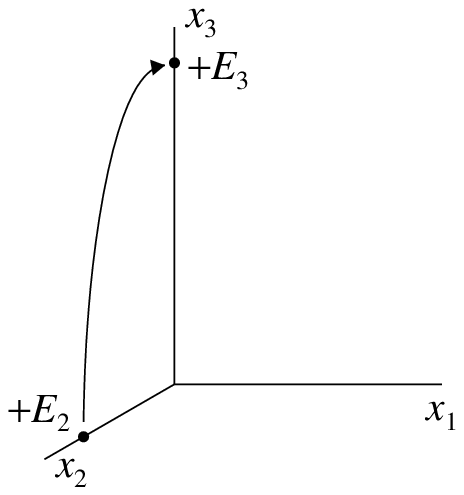}  
             \hspace{0.03\hsize}  
             \epsfxsize=0.3\hsize\epsfbox{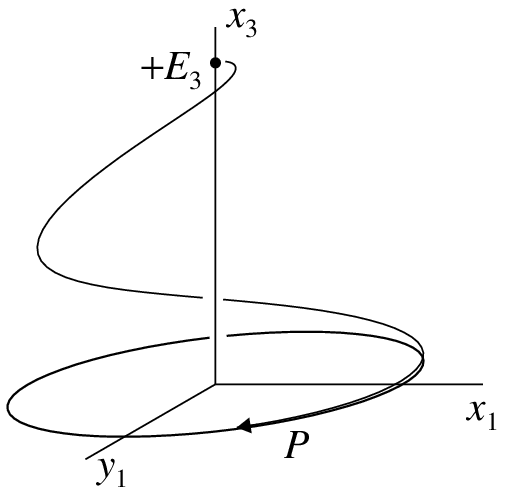}}  
 \caption{The heteroclinic cycle for the fully symmetric system.  
 (a)~One of the connections in the $x_3=0$ subspace, from the periodic orbit  
 $P$ to the equilibrium point~$+E_2$;  
 (b)~the single connection in the $x_1=y_1=0$ subspace, between the equilibria  
 $+E_2$ and~$+E_3$;  
 (c)~one of the connections in the $x_2=0$ subspace, from $+E_3$ to $P$.}  
\label{fig:cyclepicture}  
\end{figure}

We consider a system of ordinary differential equations $\dot{\mathbf  
x}={\mathbf f}({\mathbf x})$ where ${\mathbf f}:\Rset^4\to\Rset^4$, and  
${\mathbf x}=(x_1,y_1,x_2,x_3)\in\Rset^4$. It is sometimes convenient to use  
polar coordinates $(r_1,\theta_1)$ such that $z_1\equiv x_1+{\rm i}\, y_1\equiv  
r_1{\rm e}^{{\rm i}\theta_1}$. Initially, we assume the system is equivariant  
with respect to the action of a rotation and two reflections:  
$\kappa_i({\mathbf f}({\mathbf x}))={\mathbf f}(\kappa_i({\mathbf x}))$,  
$i=1,2,3$, where  
 \begin{alignat*}{2}  
 \kappa_1&\colon(z_1,x_2,x_3) & \to &(z_1{\rm e}^{{\rm i}\phi},x_2,x_3), \\  
 \kappa_2&\colon(z_1,x_2,x_3) & \to &(z_1,-x_2,x_3), \\  
 \kappa_3&\colon(z_1,x_2,x_3) & \to &(z_1,x_2,-x_3),  
 \end{alignat*}  
with $0\leq\phi<2\pi$.  
These symmetries generate the group $S^1\times Z_2\times Z_2$, and their 
presence ensures the existence  
of some dynamically invariant subspaces. We make the following assumptions  
about the dynamics in the subspaces, as illustrated in  
figure~\ref{fig:cyclepicture}:  
 \begin{itemize}  
 \item There exists a hyperbolic periodic orbit $P$ in the invariant  
plane $x_2=x_3=0$. Within this plane, the periodic orbit is a sink.  
 \item There exist hyperbolic, symmetry-related pairs of equilibria  
$\pm E_2$ and $\pm E_3$ on the invariant lines $z_1=0$, $x_3=0$ and $z_1=0$,  
$x_2=0$ respectively. Within these lines, the equilibria are sinks.  
 \item Within the invariant subspace $x_3=0$, $P$ is a saddle and  
$\pm E_2$ are sinks, and there are two-dimensional manifolds of heteroclinic  
connections from $P$ to $\pm E_2$ (figure~\ref{fig:cyclepicture}a).  
 \item Within the invariant subspace $z_1=0$, $\pm E_2$ are saddles  
and $\pm E_3$ are sinks, and there are one-dimensional heteroclinic  
connections from $+E_2$ to~$\pm E_3$, and from $-E_2$ to~$\pm E_3$  
(figure~\ref{fig:cyclepicture}b).  
 \item Within the invariant subspace $x_2=0$, $\pm E_3$ are saddles  
and $P$ is a sink, and there are two-dimensional manifolds of heteroclinic  
connections from $\pm E_3$ to $P$ (figure~\ref{fig:cyclepicture}c).  
 \end{itemize}  
  
 %%%%%%%%%%%%%%%%  
 %%%%%%%%%%%%%%%%%%%%  
  
In the presence of the rotation symmetry $\kappa_1$, the coordinate $\theta_1$  
decouples from the other coordinates, leaving an equivalent three-dimensional  
system containing a SSHC connecting three saddle-type equilibria. This cycle  
may be asymptotically stable, depending on the eigenvalues at the three  
equilibria~\cite{KrMe95}. The behaviour of trajectories near such a  
heteroclinic cycle is well understood, with a typical orbit passing near each  
of the equilibria in a cyclic manner, spending ever increasing periods of time  
near each equilibrium. The dynamics in the fully symmetric, four-dimensional  
problem therefore has analogous behaviour: trajectories cycle between two  
equilibria and a periodic orbit, with the time spent near each equilibrium or  
the periodic orbit increasing with each subsequent traversal of the  
cycle~\cite{MPR}. Moving to four dimensions does more than replace one 
pair of equilibria  
by a periodic orbit: it also introduces dynamical features that  
will be important once symmetry is broken. In particular, as can be seen in  
figure~\ref{fig:cyclepicture}, $\pm E_2$ and $\pm E_3$ are saddle-foci in the  
four-dimensional problem, and $P$ and $\pm E_3$ have two-dimensional unstable manifolds.  
  
A detailed analysis of the effect on the dynamics of small symmetry breaking is  
performed in the following sections; here we describe some geometric effects.  
Since $\pm E_2$, $\pm E_3$ and $P$ are assumed to be hyperbolic in the fully  
symmetric case, they persist and are hyperbolic when sufficiently small  
symmetry breaking terms are added. However, $+E_2$ and $-E_2$ will generically  
move off the $x_2$-axis and will no longer be related to each other by  
symmetry. Generic symmetry breaking will have an analogous effect on $+E_3$ and  
$-E_3$, and will also break the circular symmetry of $P$ and move it off  
the plane $x_2=x_3=0$.  
  
Sufficiently small symmetry breaking will not change the dimensions of the  
stable and unstable manifolds of $\pm E_2$, $\pm E_3$ and $P$, but it will  
destroy the invariant subspaces, and the heteroclinic connections that  
existed in the subspaces will either cease to exist or change their nature. We  
consider the geometric effect of symmetry breaking on each of the former  
heteroclinic connections in turn.  
  
The heteroclinic connections from $\pm E_2$ to $\pm E_3$ require the 
coincidence in $\Rset^4$ of one-dimensional and two-dimensional manifolds; 
these connections will be destroyed by a generic symmetry-breaking 
perturbation. 
  
The heteroclinic connections from $\pm E_3$ to $P$ occur when the 
two-dimensional unstable manifolds of $\pm E_3$ intersect the three-dimensional 
stable manifold of $P$. Depending on the perturbation, we generically expect to 
see either transversal intersections between these manifolds (in which case 
there are, for example, at least two robust heteroclinic connections from 
$+E_3$ to $P$) or no intersections of the manifolds. The special case where the 
manifolds are tangent can also occur in a codimension-one way. In the case of 
transversal intersections of manifolds, we might expect to see heteroclinic 
tangles and the associated complicated dynamics, depending on whether the 
dynamics elsewhere in the phase space permits reinjection of trajectories into 
the neighbourhood of the transversal intersections. 
  
The heteroclinic connections from $P$ to $\pm E_2$ occur when the 
two-dimensional unstable manifold of $P$ intersects the three-dimensional 
stable manifolds of $\pm E_2$. There is a clear analogy with the case of 
connections from $\pm E_3$ to $P$ and the comments about that case apply 
equally here. 
  
While small symmetry-breaking terms generically destroy the heteroclinic  
cycle, there will still be an attractor lying close to the original  
heteroclinic cycle (Melbourne~\cite{Me89} shows this in a closely related  
case). We show below that the form of this attractor (e.g., periodic,  
quasiperiodic, chaotic) depends on the nature of the symmetry-breaking  
perturbations included. In the fully symmetric case, the invariant  
subspaces defined by $x_2=0$ and by $x_3=0$  
restrict each trajectory to one quarter of the phase space, but 
once the reflection symmetries are  
broken, a single trajectory may explore more of the phase  
space. We are interested in determining the circumstances under which  
trajectories exhibit switching, i.e., make passages near two or more quarters  
of the original heteroclinic attractor.  
 
\section{Construction of return maps}  
 \label{sec:construct}  
  
We construct and analyse a return  
map that approximates the dynamics near the cycle. The idea is to define local coordinates  
and cross-sections near $\pm E_2$, $\pm E_3$ and $P$, then determine local maps  
valid in a neighbourhood of each of $\pm E_2$, $\pm E_3$ and $P$, and global  
maps valid in a neighbourhood of each heteroclinic connection. Composing the  
local and global maps yields the desired return map. Different forms for the return 
map are obtained depending on which of the symmetries are broken. In this section  
we list the different cases, but details of map construction are left to the Appendix. 
The techniques used are, for the most part, standard, although modifications are 
required to allow for the existence of continua of heteroclinic connections along some 
parts of the cycle in the fully symmetric case. 
 
Throughout, we use a small  
parameter~$h$ to control the size of the local neighbourhoods ($h\ll1$), and  
small parameters~$\epsilon_1$, $\epsilon_2$, $\epsilon_3$ to control the extent  
to which the symmetries $\kappa_1$, $\kappa_2$, $\kappa_3$ are broken.  
It turns out to be convenient to define the return map on a cross-section near $+E_3$. 
Using local coordinates $(r_1, \theta_1, x_2, \xi_3)$ near $+E_3$, where coordinates 
are chosen so that $+E_3$ is at the origin and so that the eigenvectors of the linearised 
flow align with the coordinate axes in the manner described in the Appendix, we define 
a cross-section  
\begin{equation} 
 H_3^\smallin  = \left\{(r_1, \theta_1,x_2,\xi_3): 0\leq r_1\leq h,  
                                           |x_2|=h,  
                                           |\xi_3|\leq h \right\}  \nonumber 
\end{equation}  
and then compute the return map, $R :   H_3^\smallin \to  H_3^\smallin$. The same 
cross-section works equally well near $-E_3$ and the maps $R$ we compute in fact approximate  
the dynamics near any of the four possible paths from $\pm E_3$ to $\pm E_3$. See the 
Appendix for details. 
 
Since we are interested in trajectories that switch between positive and 
negative values of $x_2$ and $x_3$, we introduce the notation $\pm_2$ and 
$\pm_3$ to indicate whether a trajectory visits $+E_2$ or $-E_2$, and $+E_3$ or 
$-E_3$. In particular, the trajectory starts at one of four possible sections 
specified by $H_3^\smallin$, and we use $\pm_2$ to specify whether $x_2=+h$ or 
$x_2=-h$ (implying that the trajectory recently visited $+E_2$ or $-E_2$). We 
use $\pm_3$ to specify whether the trajectory is close to $+E_3$ or $-E_3$. 
When the trajectory next returns to $H_3^\smallin$, we will be interested in 
whether it visited $+E_2$ or $-E_2$ en route, and whether it returns to $+E_3$ 
or~$-E_3$. 
  
First, in the case with full symmetry ($\epsilon_1=\epsilon_2=\epsilon_3=0$),  
we have:  
 \begin{align}  
 R(r_1,\theta_1,x_2=\pm_2h,\xi_3)=\Big(&  
           \tilde{r}_1=A r_1^{\delta},  
          \tilde{\theta}_1 = \theta_1+\Phi-Q \ln r_1, \nonumber \\  
          &\tilde{x}_2=x_2, \tilde{\xi}_3=B_2 \Big),  
 \label{eq:R_nosb}  
 \end{align}  
 where $A>0$ and $\Phi$ are constants, $\delta=\delta_1\delta_2\delta_3$, 
$Q=(e_1e_2+e_2c_3+c_3c_1)/e_1e_2e_3$, and the constants $\delta_i$, $e_i$ and $c_i$ 
are defined in the Appendix. If $x_3>0$ initially, the trajectory 
returns to $+E_3$ after visiting $\pm_2E_2$; if $x_3<0$ initially, the 
trajectory returns to~$-E_3$. 
 
Second, breaking the $\kappa_2$ and $\kappa_3$ symmetries ($\epsilon_1=0$,  
$\epsilon_2\ne 0$, $\epsilon_3\neq0$) we have:  
 \begin{align}  
 R(r_1,\theta_1,x_2=\pm_2h,\xi_3)=\Big(&  
           \tilde{r}_1=A_2\Big| \epsilon_3 \pm_3 A_1  
                          \big| \epsilon_2 \pm_2 A_3 r_1^{\delta_3}  
                                \big|^{\delta_1}  
                                \Big|^{\delta_2}, \nonumber \\  
           &\tilde{\theta}_1 = \theta_1 + \Phi_1+\Phi_2+\Phi_3  
                   -\frac{1}{e_3}\ln r_1\nonumber\\  
           &\phantom{\tilde{\theta}_1 = {}}{}  
                   -\frac{1}{e_1}\ln\big|\epsilon_2\pm_2A_3r_1^{\delta_3}\big|\nonumber\\  
           &\phantom{\tilde{\theta}_1 = {}}{}  
                   -\frac{1}{e_2}\ln\Big| \epsilon_3 \pm_3 A_1    
                          \big| \epsilon_2 \pm_2 A_3 r_1^{\delta_3}  
                                \big|^{\delta_1}  
                                \Big|,\nonumber \\  
          & \tilde{x}_2=\sgn\left(\epsilon_2\pm_2A_3r_1^{\delta_3}\right)h,  
            \tilde{\xi}_3= B_2 \Big).  
 \label{eq:R_sb23}  
 \end{align}  
The trajectory visits $+E_2$ or $-E_2$ en route according to the sign of 
$\epsilon_2\pm_2A_3r_1^{\delta_3}$, and it returns to $+E_3$ or $-E_3$ 
according to the sign of 
$\epsilon_3\pm_3A_1\big|\epsilon_2\pm_2A_3r_1^{\delta_3}\big|^{\delta_1}$. 
 
Third, if we break the $\kappa_1$ symmetry but preserve $\kappa_2$  
and~$\kappa_3$ ($\epsilon_1\neq0$, $\epsilon_2=\epsilon_3=0$) we have:  
 \begin{align}  
 R(r_1,\theta_1,x_2=\pm_2h,\xi_3)=\Big(&  
           \tilde{x}_1=\epsilon_1 a_\smallr+A_2|\hat{x}_3|^{\delta_2}  
                \cos\hat{\theta}_1,  
           \nonumber \\  
           &\tilde{y}_1=\epsilon_1 a_\smalli+A_2|\hat{x}_3|^{\delta_2}  
                \sin\hat{\theta}_1,  
           \nonumber\\  
          &\tilde{x}_2 = \sgn(\hat{x}_2)h,  
           \tilde{\xi}_3= B_2 \Big),  
 \label{eq:R_sb1}  
 \end{align}  
where  
 \begin{align*}  
 \hat{x}_2&=\pm_2\left(A_3 +  
                       \epsilon_1f_3\left(\theta_1-\frac{1}{e_3}\ln r_1\right)  
                 \right)r_1^{\delta_3},\\  
 \hat{x}_3&=\pm_3\left(A_1 +  
                       \epsilon_1f_1\left(\theta_1+\Phi_3-\frac{1}{e_3}\ln r_1  
                                                         -\frac{1}{e_1}\ln|\hat{x}_2|  
                                    \right)  
                 \right)|\hat{x}_2|^{\delta_1},\\  
 \hat{\theta}_1&=\theta_1+\Phi_1+\Phi_2+\Phi_3  
                 -\frac{1}{e_3}\ln r_1  
                 -\frac{1}{e_1}\ln|\hat{x}_2|  
                 -\frac{1}{e_2}\ln|\hat{x}_3|\,.  
 \end{align*}   
The trajectory visits $+E_2$ or $-E_2$ en route according to the sign
of~$\hat{x}_2$, and it returns to $+E_3$ or $-E_3$ according to the sign
of~$\hat{x}_3$. In this case, these signs are the same as the signs of $x_2$
and~$x_3$.
The map~(\ref{eq:R_sb1}) can be  
simplified by assuming that $A_3$ and $A_1$ are order one and dropping the  
terms proportional to $\epsilon_1$ in the expressions for $\hat{x}_2$  
and~$\hat{x}_3$. This results in an approximate map:  
 \begin{align}  
 R(r_1,\theta_1,x_2=\pm_2h,\xi_3)=\Big(&  
           \tilde{x}_1=\epsilon_1 a_\smallr+Ar_1^{\delta}  
                \cos\left(\theta_1+\Phi - Q\ln r_1\right),  
           \nonumber \\  
           &\tilde{y}_1=\epsilon_1 a_\smalli+Ar_1^{\delta}  
                \sin\left(\theta_1+\Phi - Q\ln r_1\right),  
           \nonumber\\  
          &\tilde{x}_2 = x_2,  
           \tilde{\xi}_3= B_2 \Big),  
 \label{eq:R_sb1simple}  
 \end{align}  
 where $\delta$ and $Q$ were defined above, and $A$ and $\Phi$ are constants as  
in equation~(\ref{eq:R_nosb}).  
  
Finally, when all symmetries are broken the return map is similar to the  
map~(\ref{eq:R_sb1}) above, though with definitions of $\hat{x}_2$ and  
$\hat{x}_3$ that include terms proportional to $\epsilon_2$ and $\epsilon_3$:  
 \begin{align}  
 \hat{x}_2&=\pm_2\left(A_3 +  
                       \epsilon_1f_3\left(\theta_1-\frac{1}{e_3}\ln r_1\right)  
                 \right)r_1^{\delta_3} \nonumber \\ 
          &\phantom{=}{}+\epsilon_2\left(1+\epsilon_1g_3\left(\theta_1-\frac{1}{e_3}\ln r_1\right)  
                      \right), \nonumber \\  
 \hat{x}_3&=\pm_3\left(A_1 +  
                       \epsilon_1f_1\left(\theta_1+\Phi_3-\frac{1}{e_3}\ln r_1  
                                                         -\frac{1}{e_1}\ln|\hat{x}_2|  
                                    \right)  
                 \right)|\hat{x}_2|^{\delta_1}\nonumber \\  
          &\phantom{=}{}+\epsilon_3\left(1 +  
                            \epsilon_1g_1\left(\theta_1+\Phi_3-\frac{1}{e_3}\ln r_1  
                                               -\frac{1}{e_1}\ln|\hat{x}_2|  
                                         \right)  
                      \right), \nonumber \\  
 \hat{\theta}_1&=\theta_1+\Phi_1+\Phi_2+\Phi_3  
                 -\frac{1}{e_3}\ln r_1  
                 -\frac{1}{e_1}\ln|\hat{x}_2|  
                 -\frac{1}{e_2}\ln|\hat{x}_3|\,.  
 \label{eq:R_sb1complex}  
 \end{align}  
 The trajectory visits $+E_2$ or $-E_2$ en route according to the sign 
of~$\hat{x}_2$, and it returns to $+E_3$ or $-E_3$ according to the sign 
of~$\hat{x}_3$. It might seem that terms proportional to~$\epsilon_1$ in 
$\hat{x}_2$ and $\hat{x}_3$ could be dropped, as they were above. However, the 
terms $\pm_2A_3r_1^{\delta_3}$ and $\epsilon_2$ could nearly cancel and 
likewise $\pm_3A_1|\hat{x}_2|^{\delta_1}$ and $\epsilon_3$, so we do not drop 
the $\epsilon_1$~terms. In fact, it turns out that retaining the $\epsilon_1$ 
terms is essential for understanding the switching mechanisms. 
  
It is possible to write down equivalent maps from $H_1^\smallin\to  
H_1^\smallin$ and $H_2^\smallin\to H_2^\smallin$. Note that the radial  
coordinates (as defined in the Appendix)  
play no role in the return maps, at the order to which we are  
working.

\section{Analysis of return maps}  
\label{sec:analysis} 
 
Behaviour in the case without symmetry breaking is well understood and simple:  
whenever $\delta>1$ and $r$~is small, iteration of map (\ref{eq:R_nosb})  
results in progressively smaller values of~$r$ and so there is an  
asymptotically stable heteroclinic cycle. The signs of $x_2$ and $x_3$ cannot  
change, owing to the presence of invariant subspaces, so each trajectory is  
confined to one quarter of the phase space. For the remainder of this section,  
we will assume~$\delta>1$.  
  
\subsection{Global bifurcations}  
\label{sec:global}  
  
Global bifurcations are a key ingredient for understanding the dynamics of the  
non-symmetric system. In this section, we describe the global bifurcations that  
are most important for our analysis.  
  
\subsubsection{Homoclinic bifurcation of $P$}  
\label{homP}  
  
The periodic orbit $P$ has stable and unstable manifolds of dimension three and  
two, respectively, meaning that transverse intersections of the manifolds, when  
they occur, do so in a codimension-zero way, while tangencies between the  
manifolds will be of codimension one. Transverse homoclinic orbits can only  
occur when all symmetries are broken, as the following argument shows. If  
$\epsilon_2=0$, the subspace $x_2=0$ is invariant; since $ {\cal W}^\smalls(P)$ lies in that  
subspace it cannot intersect $ {\cal W}^\smallu(P)$. Similarly, if $\epsilon_3=0$, the  
subspace $x_3=0$ is invariant; since $ {\cal W}^\smallu(P)$ lies in that subspace it cannot  
intersect $ {\cal W}^\smalls(P)$. If $\epsilon_1=0$ then the rotation symmetry ensures that  
any intersection of $ {\cal W}^\smallu(P)$ and $ {\cal W}^\smalls(P)$ will not be transverse.  
  
\begin{figure}  
 \centerline{\epsfxsize=0.5\hsize\epsfbox{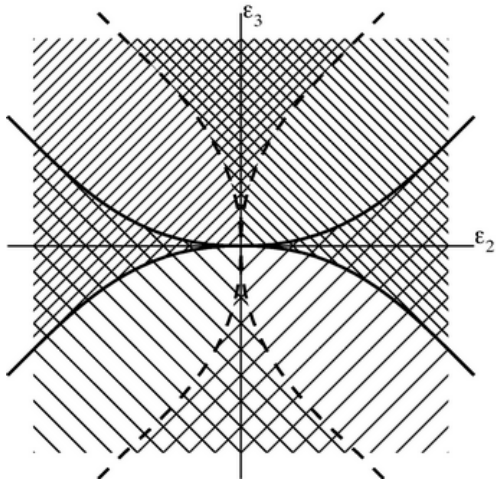}}  
\caption{Schematic bifurcation set for the case $\epsilon_1=0$, $\epsilon_2$  
and $\epsilon_3$ small. Regions of asymptotically stable quasiperiodic  
solutions are bounded by codimension-one curves of global bifurcations, i.e.,  
non-trans\-verse homoclinic bifurcations of $P$ (dashed curves) and  
heteroclinic bifurcations of the cycles $\pm E_2 \to \pm E_3 \to \pm E_2$  
(solid curves). The shapes of the global bifurcation curves correspond to the  
choice $\delta_1>1$, $\delta_2>1$ and $\delta_3>1$, but similar figures could  
be drawn for the other cases. As explained in Section~\ref{sec:reflect}, the  
various shading styles indicate the regions in which four different  
quasiperiodic solutions occur. Close to the $\epsilon_2$ and $\epsilon_3$ axes,  
two different quasiperiodic solutions coexist.}  
 \label{fig:globalbifns}  
 \end{figure}

In the case $\epsilon_1=0$, $\epsilon_2\ne 0$, $\epsilon_3 \ne 0$,  
non-trans\-versal homoclinic orbits of $P$ occur when one branch of the stable  
manifold of $P$ is coincident with one branch of the unstable manifold of $P$.  
This event can be located by calculating the image of  
$ {\cal W}^\smallu(P)$ under $\Psi_{31}\circ \phi_3 \circ \Psi_{23} \circ \phi_2 \circ \Psi_{12}$  
(see Appendix for definitions of the maps $\phi_i$ and $\Psi_{ij}$)  
and setting the $x_2$ component  
of the image to zero; we find that for small symmetry-breaking,  
non-trans\-versal homoclinic bifurcations of $P$ occur at  
 \begin{equation}  
 \epsilon_2 = -\pm_2 A_3 A_2^{\delta_3}|\epsilon_3|^{\delta_2 \delta_3},  
    \qquad \epsilon_1=0.  
    \label{eq:homP}  
 \end{equation}  
Homoclinic orbits can be formed by coincidence of either of the two branches  
of $ {\cal W}^\smallu(P)$ with either of the two branches of $ {\cal W}^\smalls(P)$, resulting in four  
possible homoclinic bifurcations corresponding to the four separate curves  
implicit in the expression above. These curves are shown as dashed lines in  
figure~\ref{fig:globalbifns}.  
The homoclinic orbit corresponding to the curve in the second quadrant of the
$(\epsilon_2,\epsilon_3)$ plane arises from the choice $\pm_2=+$ and
$\epsilon_3>0$, and passes close to $+E_2$ and $+E_3$; the three other
bifurcation curves correspond to homoclinic orbits with the three other routes
past the equilibria, in the obvious way.

 %%%%%%%%%%%%%  
 %%%%%%%%%%%%%  
  
\begin{figure}  
\centerline{\epsfxsize=0.5\hsize\epsfbox{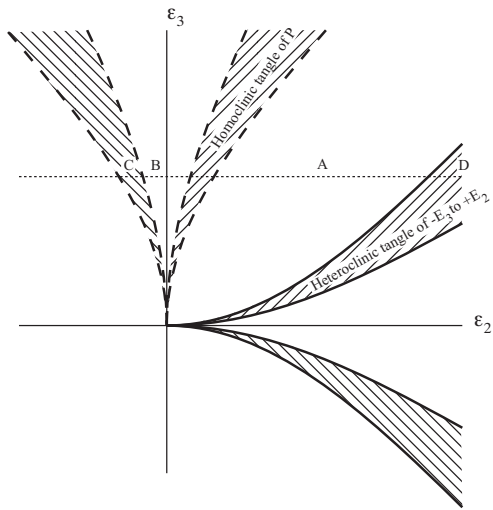}}  
\caption{Schematic diagram showing part of the bifurcation set for the case  
$\epsilon_1$ fixed and non-zero but small (compare with  
figure~\ref{fig:globalbifns}). Dashed curves correspond to homoclinic  
tangencies of $P$, solid curves in the first (resp.~fourth) quadrant correspond  
to heteroclinic tangencies between $ {\cal W}^\smallu(-E_3)$ (resp.~${\cal W}^\smallu(+E_3)$) and  
${\cal W}^\smalls(+E_2)$, and the shading shows regions in which the corresponding 
homoclinic or heteroclinic  
tangles exist. The dotted horizontal line indicates a path through parameter space  
discussed in Section~\ref{sec:full}; the labels A -- D indicate schematically 
parameter 
values used in section~\ref{sec:numerics}.}  
 \label{fig:globalbifns-sb}  
 \end{figure}

As $\epsilon_1$ changes from zero, each curve of  
non-trans\-versal homoclinic bifurcations will generically split into two  
curves of homoclinic tangencies, with the region between the tangencies being  
parameter values for which there are transverse homoclinic orbits of~$P$. Four  
curves of homoclinic tangencies and two regions of homoclinic tangles are shown  
schematically in figure~\ref{fig:globalbifns-sb}.  
  
Inspection of the expression for the $x_2$~component of the image of  
$ {\cal W}^\smallu(P)$ under $\Psi_{31}\circ \phi_3 \circ \Psi_{23} \circ \phi_2 \circ \Psi_{12}$ 
gives more information about loci  
of the homoclinic bifurcations of~$P$ when $\epsilon_1\neq0$. This component  
can be written as:  
 \begin{equation}  
 \tilde{x}_2=\pm_2 R_1^{\delta_3} \left(A_3 + \epsilon_1 f_3(\Theta_1)\right)  
             + \epsilon_2 \left(1+\epsilon_1 g_3(\Theta_1)\right),  
 \label{eq:thresholdx2}  
 \end{equation}  
where $R_1$ and $\Theta_1$ are complicated functions of the coefficients and  
parameters. In this expression, $A_3 + \epsilon_1 f_3(\Theta_1)$ must remain  
positive, as explained in the Appendix, and $R_1$ is positive. Expressions for the positions  
of the homoclinic tangencies in parameter space can be calculated by setting  
$\tilde{x}_2=0$; these expressions are not included here due to their extreme  
ugliness. Nonetheless, we note that for $\epsilon_1$ small, when $\pm_2=+$,  
there are only solutions with~$\epsilon_2<0$; this is consistent with figures  
\ref{fig:globalbifns} and \ref{fig:globalbifns-sb}, in which each bifurcation  
curve is confined to a single quadrant. However, if $\epsilon_1$ is large  
enough that $1+\epsilon_1 g_3(\Theta_1)$ can change sign as $\Theta_1$ varies,  
the loci of the homoclinic bifurcations of $P$ can change quadrants. Of course,  
this effect is outside the range of validity of the return maps we have  
constructed, but the principle is worth bearing in mind as it appears to  
influence the dynamics observed in the numerical example discussed in  
Section~\ref{sec:numerics}.  
  
 %%%%%%%%%%%%%  
 %%%%%%%%%%%%%  

\subsubsection{Heteroclinic bifurcation $\pm E_2 \to \pm E_3 \to \pm E_2$}  
\label{hetEs}  
  
In the case that all symmetries are broken, consideration of the dimensions of  
the stable and unstable manifolds of the equilibrium points shows that the  
heteroclinic cycle $+E_2 \to+E_3 \to+E_2$ will occur in a codimension-two  
manner. However, if $\epsilon_1=0$, the connection $+E_2\to+E_3$ is robust and  
the intersection of ${\cal W}^\smallu(+E_3)$ and ${\cal W}^\smalls(+E_2)$ is a codimension-one phenomenon,  
meaning that the heteroclinic cycle as a whole occurs with codimension one.  
This latter case is of interest since, as we will see, the heteroclinic  
bifurcation unfolds when $\epsilon_1 \ne 0$ into homoclinic bifurcations of  
$+E_2$ and $+E_3$ and heteroclinic tangencies between ${\cal W}^\smallu(+E_3)$  
and ${\cal W}^\smalls(+E_2)$  
similar to the way each non-trans\-verse homoclinic bifurcation of $P$ splits  
into two homoclinic tangencies when $\epsilon_1$ is varied from zero (see  
above).  An analogous argument works for heteroclinic cycles involving $-E_2$  
and/or $-E_3$.  
  
Calculations with the local and global maps yields an expression for the  
parameter values at which these heteroclinic bifurcations occur:  
 $$ \epsilon_3 = - \pm_3A_1|\epsilon_2|^{\delta_1},  
    \qquad\epsilon_1=0.  
 $$  
See figure~\ref{fig:globalbifns}. This expression is valid for all four cycles  
$\pm E_2 \to \pm E_3 \to \pm E_2$ so long as $\pm_3$ and the sign of  
$\epsilon_2$ are chosen appropriately.  
  
\subsubsection{Homoclinic bifurcations of $\pm E_2$ and $\pm E_3$}  
  
The dimensions of the stable and unstable manifolds of $\pm E_2$ and $\pm E_3$ 
are such that if homoclinic bifurcations of these equilibria occur, they are of 
codimension one. 
  
An argument similar to that used in subsection~\ref{homP} shows that we require  
$\epsilon_1 \ne 0$ and $\epsilon_3 \ne 0$ if a homoclinic bifurcation of $\pm E_2$  
is to occur, although $\epsilon_2$ could be zero. Similarly, existence of a  
homoclinic bifurcation of $\pm E_3$ requires $\epsilon_1\ne0$ and $\epsilon_2\ne0$,  
although $\epsilon_3$ could be zero. The homoclinic bifurcations of $\pm E_2$  
(resp.~$\pm E_3$) will be of Shil'nikov type if $\delta_2<1$ (resp.~$\delta_3 <1$)  
and if $c_2 < 2$ (resp.~$c_3 <2$)~\cite{GS84}.  
  
We can in principle calculate parameter values at which these homoclinic  
bifurcations occur, but the expressions are too nasty to be useful. Instead, we  
note that there can be two homoclinic bifurcations of $+E_2$, one for each  
branch of the unstable manifold of $+E_2$, and a further two homoclinic  
bifurcations of $-E_2$. Similarly, there can be two homoclinic bifurcations of  
$+E_3$ and two homoclinic bifurcations of $-E_3$. These eight homoclinic  
bifurcations will in general occur at different parameter values, but in the  
limit $\epsilon_1 \to 0$, will converge pairwise on the loci of the four  
heteroclinic bifurcations involving $\pm E_2$ and $\pm E_3$ discussed in the  
previous subsection. For instance, as $\epsilon_1 \to 0$, a homoclinic orbit of  
$+E_2$ passing near $-E_3$ and a homoclinic orbit of $-E_3$ passing near $+E_2$  
will converge in phase space on the heteroclinic cycle $+E_2\to-E_3\to+E_2$,  
and the parameter values at which the homoclinic bifurcations occur will  
converge in parameter space on the locus of the heteroclinic bifurcation. For  
clarity, these bifurcation curves are not shown in  
figure~\ref{fig:globalbifns-sb}.  
  
The dynamics associated with these bifurcations will be  
discussed further below.  
  
\subsection{Breaking the two reflection symmetries}  
 \label{sec:reflect}  
  
Here we show that, for $\epsilon_1=0$ and for sufficiently  
small~$\epsilon_2$ and~$\epsilon_3$, map~(\ref{eq:R_sb23}) generically has  
at least one asymptotically stable closed invariant curve and the corresponding  
flow has quasiperiodic solutions. This is not a surprising result, since  
the coordinate $\theta_1$ decouples from the other coordinates 
when $\epsilon_1=0$, in which case 
our system can be reduced to a three-dimensional system with a SSHC between 
equilibria; earlier work on a system related to our reduced system showed that breaking the reflection 
symmetries can give rise to asymptotically stable periodic solutions~\cite{SaSc95}. 
Our main aim in this section is to locate the regions in parameter space in which the 
quasiperiodic solutions exist, for comparison with the location of some of the global 
bifurcations described in section~\ref{sec:global}. 
  
The $r_1$ component of map~(\ref{eq:R_sb23}) is independent of the other 
variables, and so we first seek values of $r_1$ for which $F(r_1)=r_1$,  where  
 \begin{equation}  
 F(r_1)=A_2\Big| \epsilon_3 \pm_3 A_1  
                          \big| \epsilon_2 \pm_2 A_3 r_1^{\delta_3}  
                                \big|^{\delta_1}  
                                \Big|^{\delta_2}. \label{eq:f_sb23}  
 \end{equation}  
For each choice of $\epsilon_2$ and $\epsilon_3$, there are two  
possible signs of each of $\pm_2$ and~$\pm_3$, but the case $(\pm_2=+,\epsilon_2>0)$ is  
equivalent to $(\pm_2=-,\epsilon_2<0)$, and the case $(\pm_3=+,\epsilon_3>0)$  
is equivalent to $(\pm_3=-,\epsilon_3<0)$. Without loss of generality, we  
focus on the case $\pm_2=+$, $\pm_3=+$, and seek values of  
$\epsilon_2$ and $\epsilon_3$ for which there exist fixed points of  
map (\ref{eq:f_sb23}). Fixed points of this type have positive values within  
the absolute value signs in  
(\ref{eq:f_sb23}), since the signs of $\epsilon_2+A_3 r_1^{\delta_3}$ and  
$\epsilon_3+A_1|\epsilon_2+A_3r_1^{\delta_3}|^{\delta_1}$ determine the next  
values of $\pm_2$ and $\pm_3$.  
  
 %%%%%%%%%%%%%%  
\begin{figure}  
\centerline{\epsfxsize=0.9\hsize\epsfbox{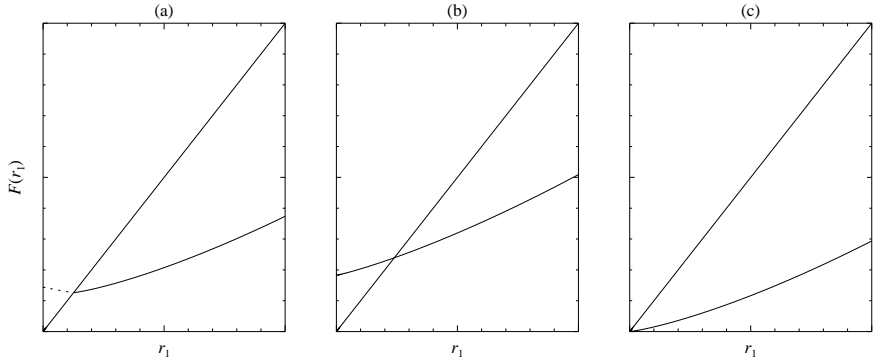}}  
\caption{Schematic graphs of $F(r_1)$ (equation (\ref{eq:f_sb23}))  
for the choice $\pm_2=\pm_3=1$ when $\delta_1>1$,  
$\delta_2>1$, $\delta_3>1$, and $A_1=A_2=A_3=1$. The solid (resp.~dotted) curve  
indicates values of $r_1$ for which the next values of $\pm_2$ and $\pm_3$ are  
(resp.~are not) both positive; we seek values of $r_1$ for which the solid  
curve intersects the diagonal. For small positive~$\epsilon_2$  
and~$\epsilon_3$, a stable fixed point exists (see panel (b)). This fixed point  
ceases to exist in the second quadrant when  
$\epsilon_2=-A_3A_2^{\delta_2}\epsilon_3^{\delta_2\delta_3}$, when there is a  
non-transversal homoclinic bifurcation of $P$ (limiting case shown in panel  
(a)). The fixed point is destroyed in the fourth quadrant when  
$\epsilon_3=-A_1\epsilon_2^{\delta_1}$, when there is a non-transversal  
heteroclinic connection from $+E_3$ to $+E_2$ (limiting case shown in  
panel~(c)).}  
 \label{fig:mapepsi23}  
 \end{figure}  
 %%%%%%%%%%%%%%  
  
For sufficiently small, positive $\epsilon_2$ and $\epsilon_3$,  
$F(0)=A_2(\epsilon_3+A_1\epsilon_2^{\delta_1})^{\delta_2}>0$. For $r_1$ larger  
than $\epsilon_2$ and $\epsilon_3$ but still smaller than one, we have  
$F(r_1)\sim r_1^{\delta}$, which is less than~$r_1$ since $\delta>1$. Thus, by  
continuity, the map has a fixed point (see figure~\ref{fig:mapepsi23}(b)).  
Since $F(r_1)$ is monotonically increasing, the slope of~$F$ at the  
fixed point is less than one, so a stable fixed point exists for 
 $\epsilon_2>0$, $\epsilon_3>0$.  
  
This fixed point (i.e., a fixed point with $\pm_2=\pm_3=+$) also exists in  
parts of the second and fourth quadrants of the $(\epsilon_2,\epsilon_3)$  
parameter plane. To determine the region of existence in the  
fourth quadrant, we fix $\epsilon_2$ at some small positive value and  
decrease $\epsilon_3$. This shifts the  
graph of $F(r_1)$ down, from which it is found that a stable fixed point  
exists until $F(0)=0$, i.e., until  
$\epsilon_3=-A_1\epsilon_2^{\delta_1}$ 
(figure~\ref{fig:mapepsi23}(c)). Thus the fixed point  
ceases to exist in the fourth quadrant at the locus of the   
heteroclinic bifurcation from $+E_3$ to $+E_2$ (c.f.~section~\ref{hetEs}).  
  
To determine where the fixed point exists in the second quadrant,  
we fix $\epsilon_3$ at some small positive value and decrease $\epsilon_2$.  
This decreases $F(0)$ and also changes the   
shape of the graph of $F(r_1)$; the graph remains monotonic  
increasing in $r_1$ while $\epsilon_2$ is positive, but develops a turning point  
once $\epsilon_2$  
becomes negative, with $F(r_1)$ decreasing for  
$r_1$ near zero. The decreasing section is indicated by a dotted curve in  
figure~\ref{fig:mapepsi23}(a), and corresponds to future values of $\pm_2$ and  
$\pm_3$ not both being positive. For small enough negative $\epsilon_2$ the  
dotted section of the graph lies to the left of fixed point, but when  
$\epsilon_2=-A_3A_2^{\delta_2}\epsilon_3^{\delta_2\delta_3}$, the dotted curve  
reaches to the diagonal and the fixed point ceases to exist  
(figure~\ref{fig:mapepsi23}(a)). Thus the  fixed point ceases to exist in the second  
quadrant on the locus of the homoclinic bifurcation of $P$ (c.f.~section \ref{homP}).  
  
The region of existence of this stable fixed point is indicated by the left-leaning  
close hatching in figure~\ref{fig:globalbifns}. Calculations with the other  
combinations of signs of $\pm_2$ and $\pm_3$ are analogous, and yield different  
regions of existence for the corresponding fixed points. Fixed points may coexist  
as shown in  
figure~\ref{fig:globalbifns}.  
  
Since for fixed $r_1$ the $\theta_1$ component of map (\ref{eq:R_sb23}) is a rigid rotation, 
a fixed point of  equation (\ref{eq:f_sb23}) generically corresponds to a closed  
invariant curve in (\ref{eq:R_sb23}) and to a quasiperiodic solution in the  
full flow.  The angle~$\theta_1$ decouples from the rest of the dynamics, and so the  
full flow will have an invariant torus foliated by periodic orbits for a dense set of  
parameter values. Stability of these solutions follows from the stability of the  
fixed point of~(\ref{eq:f_sb23}).  
  
The calculations above were for the case $\delta_1>1$, $\delta_2>1$,  
$\delta_3>1$. Similar calculations done when one or more of the  
$\delta_i$ is smaller than one lead to  
similar regions of existence of quasiperiodic solutions, except that there are 
additional saddle-node bifurcations of the tori close to the relevant  
global bifurcations; these saddle-node bifurcations arise since the  
global bifurcations destroy solutions of different stabilities  
depending on the sign of $\delta_i -1$.  
  
The special case that precisely  one of $\epsilon_2$ and $\epsilon_3$ is zero  
(i.e., only one reflection symmetry is broken) is covered by the analysis  
above; there will be two fixed points of the map with corresponding (foliated)  
tori in the flow.  
  
In summary, when the two reflection  
symmetries are broken but the rotation symmetry is preserved, the flow generically  
has asymptotically stable quasiperiodic  
solutions that do not exhibit switching. Different quasiperiodic solutions coexist in regions  
bounded by global bifurcations that will play an important role in generating complex  
dynamics when $\epsilon_1\neq0$.  
  
\subsection{Breaking the rotation symmetry}  
\label{sec:rotate}  
  
In this subsection we show that for sufficiently small values of $\epsilon_1$,  
with $\epsilon_2=\epsilon_3=0$, the map (\ref{eq:R_sb1simple}) has a stable  
fixed point.  
  
\begin{figure}  
\centerline{\epsfxsize=0.9\hsize\epsfbox{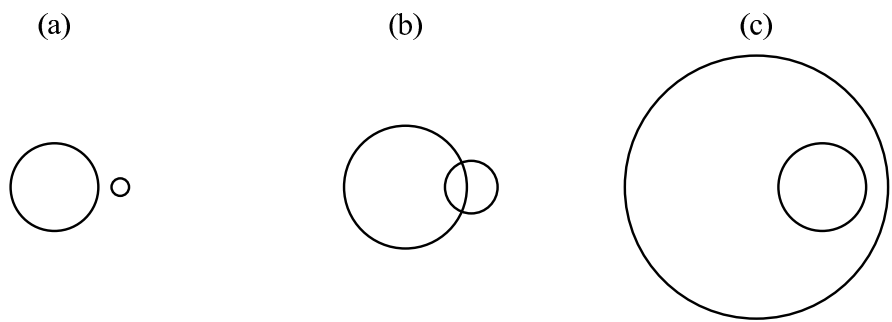}}  
\caption{Finding fixed points of equations (\ref{eq:Rnorotrhophi}). Panels  
(a--c) show schematically the relative positions of the circle of radius $r$  
around $(x_1,y_1)=(0,0)$ (large circle in each panel) and its image under map  
(\ref{eq:Rnorotrhophi}) for various sizes of $r$: (a) $r <  
\epsilon_1-r^{\delta}$, no intersection; (b) $r \geq \epsilon_1-r^{\delta}$ and  
$r \leq \epsilon_1+r^{\delta}$, one or two intersections; (c) $r >  
\epsilon_1+r^{\delta}$, no intersections.} 
 \label{fig:tildephi1}  
 \end{figure}

By rescaling $r_1$ and $\epsilon_1$ by order one amounts and moving the origin  
of the $\theta_1$ coordinate, we can without loss of generality set $A=1$,  
$a_\smallr=1$ and $a_\smalli=0$ in (\ref{eq:R_sb1simple}). Ignoring for now the  
$x_2$ and $\xi_3$ components of the map and working with polar coordinates  
$(\rho,\phi)$ centred at $(x_1,y_1)=(\epsilon_1,0)$ (so that  
$x_1=\epsilon_1+\rho\cos \phi$ and $y_1=\rho \sin \phi$), map  
(\ref{eq:R_sb1simple}) reduces to  
 \begin{equation}  
 \begin{cases}  
   \tilde{x}_1=\epsilon_1 + \tilde\rho\cos\tilde\phi\cr  
   \tilde{y}_1=\phantom{\epsilon_1+{}} \tilde\rho\sin\tilde\phi\cr  
 \end{cases}  
 \qquad \mbox{where} \qquad  
 \begin{cases}  
   \tilde\rho = r_1^\delta\cr  
   \tilde\phi = \theta_1+\Phi - Q\ln r_1.\cr  
 \end{cases}  
 \label{eq:Rnorotrhophi}  
 \end{equation}  
The constant $\Phi$ may take a different value here than in  
equations~(\ref{eq:R_sb1simple}). Fixed points of (\ref{eq:Rnorotrhophi})  
satisfy $r_1=\sqrt{{\tilde{x}_1}^2+{\tilde{y}_1}^2}$ and  
$\theta_1=\arctan\left(\tilde{y}_1/\tilde{x}_1\right)$. To find solutions of  
the first of these equations, note that circles of radius $r$ about  
$(x_1,y_1)=(0,0)$ map under~(\ref{eq:Rnorotrhophi}) to circles of radius  
$r^\delta$ about $(\epsilon_1,0)$. Since $\delta>1$ and for small $r$, these  
circles will intersect if $r\geq\epsilon_1-r^\delta$ and  
$r\leq\epsilon_1+r^\delta$; the intersection points are candidate fixed points  
of the map. For each small fixed value of $\epsilon_1$, there will be some  
non-zero interval $a\leq r\leq b$ on which the inequalities are both  
satisfied.  See figure \ref{fig:tildephi1}(a--c).  
  
\begin{figure}  
\centerline{\epsfxsize=0.9\hsize\epsfbox{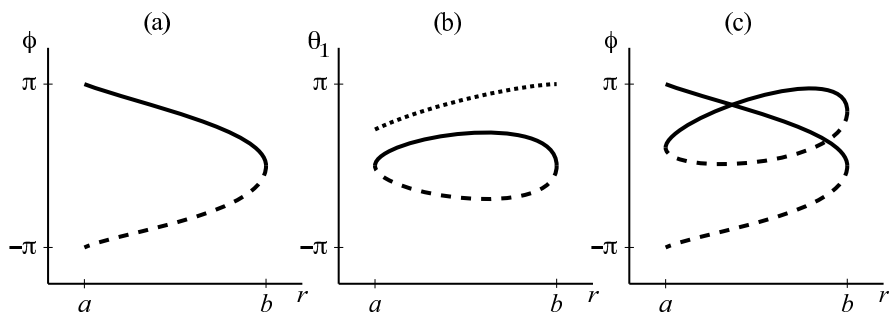}}  
\caption{(a)~Schematic $\phi$ values for intersections of the two circles 
$(r_1, \theta_1)=(r, \theta_1)$ and $(\rho,\phi)=(r^\delta, \phi)$, in the 
situation shown in figure~\ref{fig:tildephi1}(b), plotted as a function of $r$; 
 (b)~Schematic showing $\theta_1$ (solid and dashed curves) and $-Q\ln r_1$  
(dotted curve) for points of intersection of the two circles, plotted as a  
function of $r$ at the intersection points;  
 (c)~$\phi$ and $\tilde\phi$ values at points of intersection of the two  
circles. Solid curves correspond to the upper intersection  
points and their images, dashed curves correspond to the lower intersection  
points and their images.}  
 \label{fig:tildephi2}  
 \end{figure}

The second equation, $\theta_1=\arctan\left(\tilde{y}_1/\tilde{x}_1\right)$, is  
satisfied for at least one value of~$r$ in $[a,b]$, as the following argument  
shows. When $r=a$, the circles $(r_1, \theta_1)=(r, \theta_1)$ and  
$(\rho,\phi)=(r^\delta, \phi)$ intersect at a single point,  
$(r_1,\theta_1)=(a,0)$, alternatively $(\rho, \phi)=(a^\delta, \pi)$. As $r$ is  
increased beyond~$a$, the intersection point splits into two (with  
corresponding $\phi$ values just below $\pi$ and just above $-\pi$). The  
intersection points come together again at $(r_1,\theta_1)=(b,0)$ or  
$(\rho,\phi)=(b^\delta,0)$, in the manner shown in  
figure~\ref{fig:tildephi2}(a). The corresponding value of $\tilde\phi$ for each  
intersection point can be calculated from~(\ref{eq:Rnorotrhophi}) (see  
figure~\ref{fig:tildephi2}(b)) from which it is seen that the two branches of  
$\tilde\phi$, arising from the upper and lower intersections of the two  
circles, start and end at the same point as each other, as shown in  
figure~\ref{fig:tildephi2}(c). At least one of the two branches of the graph of  
$\tilde\phi$ vs $r$ therefore intersects the graph of $\phi$ vs $r$ for that  
same branch for at least one $r$ in $[a,b]$. Thus, there is at least  
one fixed point of the map~(\ref{eq:Rnorotrhophi}).  
  
 %%%%%%%%%%%%%%%%%  
 %%%%%%%%%%%%%%%%%  
  
 %%%%%%%%%%%%%%%%%  
 %%%%%%%%%%%%%%%%%  
  
It is straightforward to show that the determinant of the Jacobian of  
map~(\ref{eq:Rnorotrhophi}) is $\delta r^{2(\delta-1)}$ and that the absolute  
value of the trace of the Jacobian is bounded above by $  
\sqrt{(\delta+1)^2+Q^2}\, r^{\delta-1}.$ For each choice of $\epsilon_1$, the  
value of $r$ at the corresponding fixed point lies in $[a,b]$, where $a$ and  
$b$ depend on $\epsilon_1$ with $a>0$ and $a$ and $b$ tending to zero as  
$\epsilon_1 \to 0$. Thus, for sufficiently small $\epsilon_1$, the determinant  
and trace of the linearised map are small enough to ensure that the  
relevant fixed point is stable. It follows that the corresponding fixed point  
of equations (\ref{eq:R_sb1simple}) is also stable, since orbits of that  
map collapse onto a constant value of the third (radial) coordinate after one  
iteration of the map.  
  
In summary, for sufficiently small $\epsilon_1$ when $\epsilon_2=\epsilon_3=0$, the  
heteroclinic cycle that occurred in the fully symmetric case is replaced by a  
stable periodic orbit.  Using the reflection symmetries $\kappa_2$ and  
$\kappa_3$, we find four stable periodic orbits co-exist, one in each quarter  
of the phase space.  
  
\subsection{Breaking all symmetries}  
\label{sec:full}  
  
Direct analysis of the return map valid for the case that all symmetries are  
broken is not feasible because of the extremely complicated form of that map.  
Instead, in this section we use our knowledge of the dynamics in the case  
$\epsilon_1=0$ and arguments about generic unfoldings of this special case to  
deduce what types of dynamics will be seen in the fully asymmetric case for  
$\epsilon_1$ near zero, and to (approximately) locate each type of behaviour in  
parameter space. This procedure allows us to make specific predictions about  
the mechanisms underlying the complicated dynamics observed in numerical  
examples, such as the example described in section~\ref{sec:numerics}. We are  
especially interested in finding mechanisms that cause repeated, non-periodic  
switching in our system.  
  
The dynamics associated with the case $\epsilon_1=0$ is summarised in  
figure~\ref{fig:globalbifns}, which shows eight curves of global bifurcations  
bounding regions in which there are asymptotically stable quasiperiodic  
solutions. In this case the rotation symmetry prevents coupling of the two  
frequencies associated with each quasiperiodic solution and the dynamics is  
simple. Once $\epsilon_1$ moves away from zero, we will generically see locking  
of the frequencies. For instance, if we were to fix $\epsilon_1$ sufficiently  
small but non-zero and pick $\epsilon_2$ and $\epsilon_3$ positive and with  
values midway between the two pairs of global bifurcation curves in the first  
quadrant of figure~\ref{fig:globalbifns-sb}, then along a one-dimensional path  
through the parameter space such as  the dotted line in  
figure~\ref{fig:globalbifns-sb} there will be intervals of quasiperiodicity  
interspersed with intervals of locked, periodic behaviour. Associated with the  
frequency locking there may be complicated dynamics such as period-doubling  
cascades and chaotic dynamics. However, this behaviour will mostly be confined  
to regions of phase space near the original quasiperiodic solutions, and is not  
the main mechanism for switching in our system.  
  
As shown in section~\ref{homP} and in  
figure~\ref{fig:globalbifns-sb}, each curve of non-transverse homoclinic  
bifurcations of $P$ seen in figure~\ref{fig:globalbifns} turns into a wedge in  
parameter space of transverse homoclinic orbits of $P$  
when $\epsilon_1$ changes from zero. There will be horseshoes and chaotic  
dynamics associated with the transverse homoclinic orbits, although the chaos  
may not be attracting. In numerical examples we might expect to see a mixture  
of stable periodic orbits and stable chaotic dynamics, in overlapping regions  
of parameter space.  
  
An interesting consequence of the occurrence of homoclinic tangles is that it 
provides a mechanism for switching of orbits with respect to the $x_2$ 
variable.  For instance, for sufficiently small $\epsilon_1$, and with 
$\epsilon_2>0$, $\epsilon_3>0$ and both small, solutions that make excursions 
near $+E_2$ can get trapped. The trapping region is bounded in part by one 
branch of ${\cal W}^\smalls(P)$ and trapped solutions make excursions past 
$+E_2$ but cannot cross ${\cal W}^\smalls(P)$ so cannot get close to $-E_2$. If 
$\epsilon_2$  is decreased, say by moving along the dotted path shown in Figure 
4 into the homoclinic wedge in the second quadrant, the trapping region 
develops a leak when a homoclinic tangency forms between that branch of ${\cal 
W}^\smalls(P)$ and a branch of ${\cal W}^\smallu(P)$; solutions are then able 
to cross ${\cal W}^\smalls(P)$, and may visit a neighbourhood of $-E_2$. We 
call this `switching in $x_2$'. Switching of this type (from positive to 
negative $x_2$) can occur for parameter values to the left of the right 
boundary of the homoclinic wedge in the second quadrant. A numerical example of 
this leaking process is given below in figure~\ref{fig:x2switchp}. 
  
Once a switching solution arrives in the region with $x_2<0$, it may then get  
captured by an attractor lying solely in the negative $x_2$ region of phase  
space, in which case no more switching will be observed. Alternatively, if  
there is a mechanism for orbits to leak back to the original region of phase  
space then there could be sustained switching in $x_2$. This latter case cannot  
occur for arbitrarily small $\epsilon_1$ as the following argument shows. For  
the case $\epsilon_1=0$, results from section~\ref{sec:reflect} show that  
orbits which make repeated excursions near $-E_2$ and $+E_3$ occur in the second  
quadrant of figure~\ref{fig:globalbifns-sb} and in the first quadrant as far as  
the locus of homoclinic bifurcations of $P$. (This homoclinic bifurcation  
involves a different branch of ${\cal W}^\smallu(P)$ than the homoclinic bifurcation  
occurring for $\epsilon_2<0$ discussed in the last paragraph.) For small  
$\epsilon_1$, there will be a wedge of homoclinic tangencies of the relevant  
branches of ${\cal W}^\smallu(P)$ and ${\cal W}^\smalls(P)$, with the wedge lying entirely within the  
region $\epsilon_2>0$, $\epsilon_3>0$; orbits that make a number of excursions  
past $-E_2$ before switching and passing close to $+E_2$ can only occur for  
values of $\epsilon_2$ and $\epsilon_3$ lying to the right of the left boundary  
of the wedge. Thus, for sufficiently small $\epsilon_1$, there is no overlap  
between the region of parameter space where there is switching from positive to  
negative $x_2$ and the region where there is switching from negative to  
positive $x_2$, with the consequence that there can be no sustained switching  
in $x_2$.  
  
However, as argued in section~\ref{homP}, for large enough $\epsilon_1$ the  
curves of homoclinic tangency may change quadrants in the  
$(\epsilon_2,\epsilon_3)$ parameter plane, and then the switching regions can  
overlap, making sustained switching in $x_2$ possible. As pointed out in  
section~\ref{homP}, our return map construction is not valid for `large'  
$\epsilon_1$, so we have not proved the existence of sustained switching, just  
shown how it might feasibly occur. It is not possible to determine a priori how  
big $\epsilon_1$ would have to be to get sustained switching, but we have shown  
that there is a threshold in $\epsilon_1$ below which sustained switching in  
$x_2$ is not possible. The value of this threshold does not go to zero as  
$\epsilon_2$ and $\epsilon_3$ go to zero, as it comes from the requirement that  
$1+\epsilon_1g_3(\Theta_1)$ changes sign (as a function of~$\Theta_1$) in  
equation~(\ref{eq:thresholdx2}). Another way of understanding this is to note  
that in~(\ref{eq:thresholdx2}), if $\pm_2=+$ and $\epsilon_2>0$, then the only  
way of having $\tilde{x}_2$ negative is to have $1+\epsilon_1g_3(\Theta_1)<0$  
for some value of~$\Theta_1$. This is a necessary but not sufficient condition,  
as the attractor may not explore the required range of~$\Theta_1$.  
  
The four curves of heteroclinic bifurcations of the cycles $\pm E_2 \to \pm E_3 
\to \pm E_2$ shown in figure~\ref{fig:globalbifns} also split when $\epsilon_1$ 
becomes non-zero, being replaced by eight curves of homoclinic bifurcations and 
eight curves  of heteroclinic tangencies between ${\cal W}^\smallu( \pm E_3)$ 
and ${\cal W}^\smalls(\pm E_2)$,  as described in section~\ref{hetEs}. If they 
are of Shil'nikov type, the homoclinic bifurcations can complicate the dynamics 
by inducing chaotic dynamics. The heteroclinic bifurcations are associated with 
switching in the $x_3$ coordinate similarly to the way switching in $x_2$ is 
associated with homoclinic bifurcations of $P$, described above. More 
precisely, the eight curves of heteroclinic tangencies between ${\cal 
W}^\smallu(\pm E_3)$ and ${\cal W}^\smalls(\pm E_2)$ come in pairs, with each 
pair bounding a wedge in parameter space. At parameter values within each wedge 
there is a heteroclinic tangle of one pair of manifold branches. For instance, 
for sufficiently small $\epsilon_1$ there will be a heteroclinic wedge 
involving one branch of ${\cal W}^\smallu(+E_3)$ and one branch of ${\cal 
W}^\smalls(+E_2)$ occurring in the fourth quadrant in the 
$(\epsilon_2,\epsilon_3)$ plane. Above this wedge, ${\cal W}^\smalls(+E_2)$ 
bounds in part a trapping region; orbits in the trapping region make excursions 
past $+E_3$ but cannot cross ${\cal W}^\smalls(+E_2)$ to get close to $-E_3$. 
The trapping region develops a leak when a heteroclinic tangency forms between 
the appropriate branches of ${\cal W}^\smallu(+E_3)$ and ${\cal 
W}^\smalls(+E_2)$ thus allowing solutions to cross ${\cal W}^\smalls(+E_2)$. We 
call this `switching in $x_3$'. An argument analogous to that used for 
switching in $x_2$ can be used here to show that there is a (generically 
different) threshold in $\epsilon_1$ below which there can be no persistent 
switching in~$x_3$. This threshold does not go to zero when $\epsilon_2$ and 
$\epsilon_3$ go to zero. 
  
The mechanisms inducing switching in $x_2$ and in $x_3$ are distinct, but  
orbits that switch persistently in both $x_2$ and $x_3$ are possible for  
$\epsilon_1$ above the thresholds for both mechanisms. Switching in each  
variable requires the rotation and appropriate reflection symmetry to be  
broken. It is possible to have persistent switching in~$x_2$ with  
$\epsilon_3=0$, or switching in~$x_3$ with $\epsilon_2=0$, though we will not  
explore this possibility in detail. The example in section \ref{sec:numerics}  
shows that persistent switching is easily observed in numerical examples.  
  
Both of the global bifurcations we have identified as inducing switching, i.e.,  
homoclinic tangencies of ${\cal W}^\smallu(P)$ and ${\cal W}^\smalls(P)$ and heteroclinic tangencies of  
${\cal W}^\smallu(\pm E_3)$ and ${\cal W}^\smalls(\pm E_2)$, will produce horseshoes in the dynamics. In  
the case of homoclinic tangencies, this is a standard result and in the case of  
the heteroclinic tangencies, reinjection into the neighbourhood of the  
heteroclinic tangle is provided by proximity in phase and parameter space to  
the heteroclinic cycle $\pm E_2 \to \pm E_3 \to \pm E_2$. In either case, we  
expect the onset of switching to be commonly associated with nearby  
chaotic dynamics; chaotic orbits before the onset of switching, chaotic  
transients for switching orbits and orbits that switch chaotically might all be  
seen. However, other types of switching are also possible, such as  
periodic switching where the attractor is a periodic orbit that crosses the  
(non-invariant) hyperplanes $x_2=0$ and $x_3=0$ or periodic switching where the  
attractor is a `noisy periodic orbit' such as results from a cascade of period  
doubling. In the latter case the itinerary of visits to $\pm E_2$ or $\pm E_3$  
will be periodic even though the actual orbits are not.  
  
\section{Example}  
\label{sec:numerics}  
  
 We consider the following system of equations to illustrate the dynamics of  
interest in this paper:  
 \begin{align}  
 {\dot z_1} &= (1+{\rm i})z_1 - |z_1|^2z_1  
              - (c_2+1)x_2^2z_1  
              + (e_3-1)x_3^2z_1  
              + \epsilon_1d_{11} + \epsilon_1d_{12}x_1\,,  
  \label{eq:modelSBODEsfirst}\\  
 {\dot x_2} &= x_2 - x_2^3  
              - (c_3+1)x_3^2x_2  
              + (e_1-1)|z_1|^2x_2  
              + \epsilon_2d_{21} + \epsilon_1d_{22}x_1x_2  
              + \epsilon_1\epsilon_2d_{23}x_1\,,  
  \\  
 {\dot x_3} &= x_3 - x_3^3  
              - (c_1+1)|z_1|^2x_3  
              + (e_2-1)x_2^2x_3  
              + \epsilon_3d_{31} + \epsilon_1d_{32}x_1x_3  
              + \epsilon_1\epsilon_3d_{33}x_1\,,  
  \label{eq:modelSBODEslast}  
 \end{align}  
These equations were derived by starting with the structurally stable
heteroclinic cycle considered in~\cite{GuHo88}, turning a pair of equilibria of
that cycle into a periodic orbit by adding a trivial phase variable, and adding
the simplest possible terms that break the symmetries in generic ways. The
parameters $\epsilon_1$, $\epsilon_2$ and $\epsilon_3$ in equations
(\ref{eq:modelSBODEsfirst}--\ref{eq:modelSBODEslast}) play the same role as in
the maps derived earlier in this paper.
  
The model used in~\cite{MPR} is similar, but differs in two  
respects. First, the symmetry-breaking terms in~\cite{MPR} were fifth-order in  
the $x$ and $z$ variables, rather than constant, linear and quadratic here.  
Second, the model in~\cite{MPR} respects the symmetry  
$(z_1,x_2,x_3)\rightarrow(-z_1,-x_2,-x_3)$, as appropriate for a model of  
a dynamo instability: the invariant subspace $z_1=x_2=x_3=0$ corresponds to the  
absence of any magnetic field. We do not expect the first difference between   
models to alter the qualitative behaviour, but the enforced symmetry may  
have a significant effect, as discussed briefly in the next section.  
  
The coefficients in the equations were chosen to be:  
  $c_1=1.2$, $e_1=1.0$,  
  $c_2=1.1$, $e_2=1.0$,  
  $c_3=1.1$, $e_3=1.0$ for the contracting and expanding eigenvalues, and  
  $d_{11}=d_{12}=10^{-4}$,  
  $d_{21}=10^{-1}$,  
  $d_{22}=10^{-1}$,  
  $d_{23}=10^{3}$,  
  $d_{31}=10^{-3}$,  
  $d_{32}=10^{-4}$,  
  $d_{33}=1$ for the symmetry-breaking coefficients.  
 The eigenvalues were chosen to be of order one, with contraction dominating  
expansion at each point ($\delta_1=1.2$, $\delta_2=1.1$, $\delta_3=1.1$, and an  
overall $\delta=1.452$). The symmetry breaking coefficients are notionally  
small, but those coefficients ($d_{23}$ and $d_{33}$) that are multiplied by  
two $\epsilon$'s were chosen to be larger to compensate for this. The exact  
numbers are not important, though they will affect the details of what is   
observed. However, choosing $d_{23}$ and $d_{33}$ to be reasonably large 
means that the switching dynamics is easier to obtain for small values  
of~$\epsilon_1$: in order to get persistent switching, the  
$\epsilon_1\epsilon_2d_{23}x_1$ and $\epsilon_1\epsilon_3d_{33}x_1$ terms in 
(\ref{eq:modelSBODEsfirst}--\ref{eq:modelSBODEslast}) need to be reasonably 
important. 
  
We integrated the equations numerically using the Bulirsch--Stoer adaptive  
integrator \cite{NumRecipes}, with a tolerance for the relative error set to  
$10^{-12}$ for each step. Poincar\'e sections were computed using algorithms   
from~\cite{PC89}.  
  
By varying $\epsilon_1$, $\epsilon_2$ and $\epsilon_3$, we are able to find  
examples of the important symmetry-breaking effects discussed in the previous  
sections of this paper. The cases with full or partial symmetry preserved give  
straightforward results, which we describe only briefly; more details are  
provided of the case of fully broken symmetry.  
  
If all symmetries are preserved  (all $\epsilon_i=0$) then each solution  
starting off the invariant subspaces is attracted to one of four symmetry-related 
structurally stable heteroclinic cycles.  
If  $\epsilon_1 \ne 0$, $\epsilon_2=\epsilon_3=0$ (rotation  
symmetry broken, reflections preserved), numerics confirm the predictions of  
section~\ref{sec:rotate}, and a single attracting periodic orbit is found in  
each quarter of the phase space. If rotation symmetry is preserved as well as  
one reflection, and the other reflection is broken, then solutions are  
attracted to a foliated torus, as discussed in section~\ref{sec:reflect}. In  
the case that both reflections are broken but the rotation symmetry is  
preserved, numerics confirm the predictions of section~\ref{sec:reflect}; we  
find that there exist attracting quasiperiodic solutions in regions bounded by  
curves of global bifurcations, as shown schematically in  
figure~\ref{fig:globalbifns}. Analysis of the maps derived earlier allows us to  
predict scaling of the loci of various global bifurcations in the limit of  
small symmetry breaking. For instance,  equation (\ref{eq:homP}) tells us that  
for $\epsilon_1=0$ and $\epsilon_3 \to 0$, homoclinic bifurcations of $P$  
associated with equations (\ref{eq:modelSBODEsfirst}--\ref{eq:modelSBODEslast})  
occur for $\epsilon_2={\rm constant} \times |\epsilon_3|^{\delta_2 \delta_3}$  
but the  value of the constant is not determined by the map analysis. Numerical  
simulations of equations (\ref{eq:modelSBODEsfirst}--\ref{eq:modelSBODEslast})  
confirm the scalings for the various global bifurcations.  
  
To illustrate the phenomena associated with breaking all symmetries, it is  
helpful to consider the changes in dynamics seen along a one-dimensional path  
such as that shown as the dotted line in figure \ref{fig:globalbifns-sb}. We  
first chose a value of~$\epsilon_1$ below the thresholds for persistent  
switching in $x_2$ and~$x_3$. For instance, fixing  $\epsilon_1=10^{-4}$ and  
$\epsilon_3=0.001$ and allowing $\epsilon_2$ to vary, we see the following  
types of dynamics.  
  
Picking $\epsilon_2=3\times10^{-5}$ yields 
a point (labelled~A) lying to the right of the homoclinic wedge and to left of the  
heteroclinic wedge in the first quadrant of figure~\ref{fig:globalbifns-sb}.  
For these $\epsilon_2$ and $\epsilon_3$ values but for $\epsilon_1=0$, there  
exists an attracting quasiperiodic solution with $x_2$ and $x_3$ both positive.  
With $\epsilon_1=10^{-4}$ the same type of quasiperiodic solution exists. As  
$\epsilon_2$ is decreased while $\epsilon_1$ is fixed at $10^{-4}$  
we find, as expected, intervals of $\epsilon_2$ in  
which there are quasiperiodic attractors interspersed with intervals on which  
there is locking of the two frequencies associated with the quasiperiodic  
solution (periodic orbits). In some intervals, period doubling cascades are  
observed, as is normal near quasiperiodic behaviour.  The interchange between  
locking and quasiperiodic behaviour persists until we approach the homoclinic  
wedge at negative values of $\epsilon_2$.  
  
 %%%%%%%%%%%%%%%%  
\begin{figure}  
\centerline{\epsfxsize=0.9\hsize\epsfbox{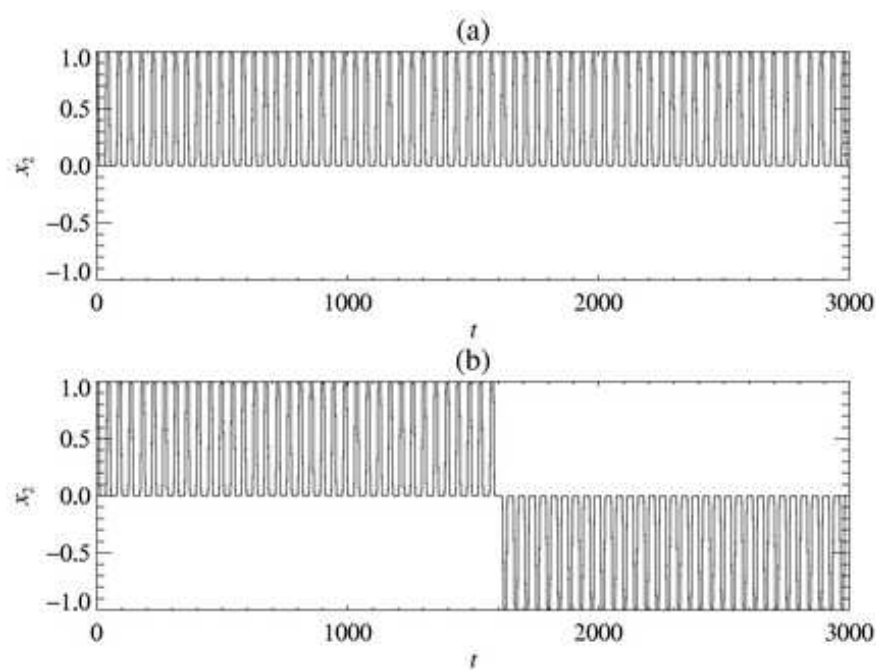}}  
 % \mbox{\includegraphics[width=0.9\hsize]{kr_fig_x2_switch.eps}}  
\caption{Onset of switching in $x_2$ for equations  
(\ref{eq:modelSBODEsfirst}--\ref{eq:modelSBODEslast}), associated with crossing  
into the wedge of homoclinic bifurcations of $P$ in the second quadrant of  
$(\epsilon_2,\epsilon_3)$-space.  Parameters $\epsilon_1=10^{-4}$,  
$\epsilon_3=0.001$ are fixed and $\epsilon_2$ is decreased: (a)  
$\epsilon_2=-5.67\times10^{-7}$ gives a chaotic attractor confined to the region  
$x_2>0$; (b) $\epsilon_2=-5.68\times10^{-7}$ gives a chaotic transient with $x_2>0$,  
then the sign of $x_2$ changes and the orbit is attracted to a quasiperiodic  
solution with $x_2<0$. Other coefficients as described in text. The  
chaotic nature of the orbit in (a) and the transient in (b) is not apparent on  
the timescale used to plot the time series.}  
 \label{fig:x2switch}  
\end{figure}  
 %%%%%%%%%%%%%%%%%  
  
 %%%%%%%%%%%%%%%%  
\begin{figure}  
\centerline{\epsfxsize=0.9\hsize\epsfbox{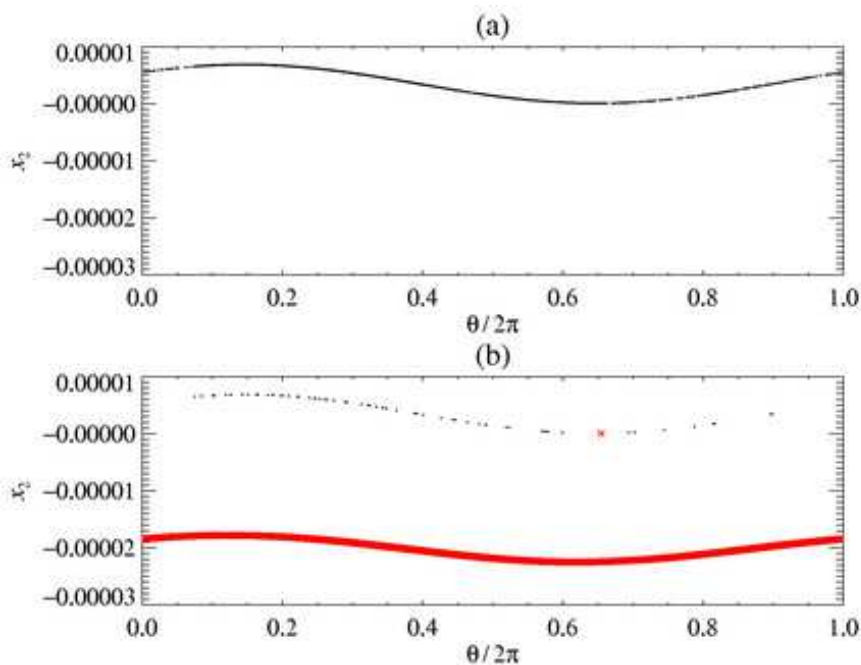}}  
 % \mbox{\includegraphics[width=0.9\hsize]{kr_fig_x2_switch_p.eps}}  
\caption{Illustration of how leaking develops as the attractor crosses 
${\cal W}^\smalls(P)$ (approximately $x_2=0$ in this figure). 
(a) Poincar\'e map for the orbit shown in  
figure~\ref{fig:x2switch}(a); (b)~Poincar\'e map for the orbit in  
figure~\ref{fig:x2switch}(b). In each case the Poincar\'e section is  
$H_1^\smallin$ with $|x_3|=h=0.005$. In each panel, a dot (resp.~cross)  
indicates that the orbit next crosses the Poincar\'e section with $x_2>0$  
(resp.~$x_2<0$). In~(b), the upper collection of dots corresponds to the   
chaotic transient, which ends in a single cross, after which the orbit switches   
to a quasiperiodic attractor, represented by the lower collection of crosses.}  
 \label{fig:x2switchp}  
\end{figure}  
 %%%%%%%%%%%%%%%%%  
  
As this homoclinic wedge is approached, apparently chaotic dynamics is  
observed, consistent with the appearance of horseshoes associated with the  
impending homoclinic tangency. Before the first tangency is reached 
(labelled~B in figure~\ref{fig:globalbifns-sb}), orbits are  
trapped in the region with  $x_2>0$, $x_3>0$ (figure~\ref{fig:x2switch}a). If  
$\epsilon_2$ is decreased past the tangency value (labelled~C), the attractor crosses  
${\cal W}^\smalls(P)$ and so a typical orbit will display a chaotic transient with $x_2>0$,  
$x_3>0$, and then switch to $x_2<0$, $x_3>0$, after which the orbit is  
attracted to a quasiperiodic solution in that quarter of phase space  
(figure~\ref{fig:x2switch}b). Corresponding Poincar\'e maps are shown in  
figure~\ref{fig:x2switchp}. With negative  
values of~$\epsilon_2$, once the trajectory has switched to $x_2<0$, the  
behaviour is analogous to that observed with $\epsilon_2>0$ and~$x_2>0$:  
quasiperiodic attractors interspersed with frequency locking. Note there is no  
persistent switching in~$x_2$, and no switching in~$x_3$, for these parameter  
values.  
  
If $\epsilon_2$ is now increased from $\epsilon_2=3\times10^{-5}$ while $\epsilon_1$  
and $\epsilon_3$ are kept fixed as before, we approach the wedge of  
heteroclinic connections from $-E_3$ to $+E_2$ (following the dotted line in  
figure~\ref{fig:globalbifns-sb}). Once the left edge of the heteroclinic wedge  
has been crossed, orbits can switch from $x_3<0$ to $x_3>0$ (but not from  
$x_3>0$ to $x_3<0$). For example, at $\epsilon_2=1.1\times10^{-4}$ (labelled~D in  
figure~\ref{fig:globalbifns-sb}), we find a chaotic  
transient with $x_3<0$ that has a single switch to a locked periodic orbit  
with~$x_3>0$, in a manner similar to the single $x_2$ switch in  
figure~\ref{fig:x2switch}(b). For larger $\epsilon_2$, we find other examples  
of quasiperiodicity, locked periodic orbits, chaos, chaotic transients and  
single switches from $x_3<0$ to $x_3>0$, consistent with the analysis presented  
above. We did not find any examples of persistent switching.  
  
\begin{figure}  
\centerline{\epsfxsize=0.9\hsize\epsfbox{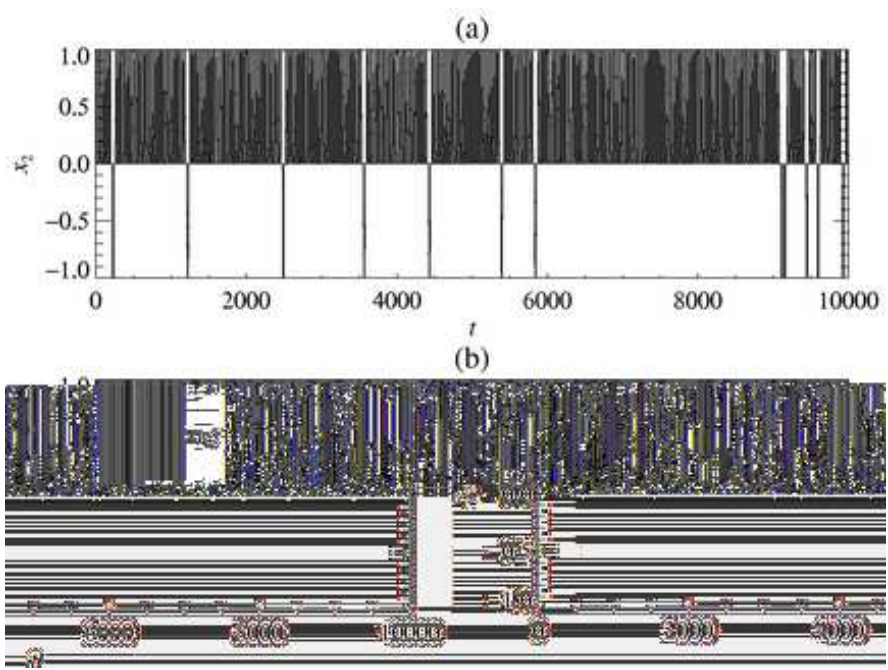}}  
 % \mbox{\includegraphics[width=0.9\hsize]{kr_fig_x2_persistent_switch.eps}}  
\caption{Time series showing persistent switching in $x_2$ alone, for equations  
(\ref{eq:modelSBODEsfirst}--\ref{eq:modelSBODEslast}) with $\epsilon_1=3\times10^{-4}$,  
$\epsilon_2=2\times10^{-4}$, $\epsilon_3=0.001$. The other coefficients are as defined  
in text.}  
 \label{fig:x2perswitch}  
\end{figure}

\begin{figure}  
\centerline{\epsfxsize=0.9\hsize\epsfbox{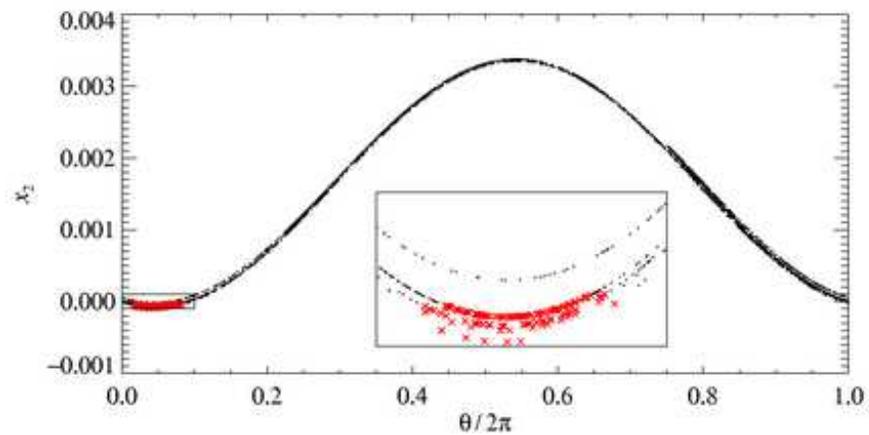}}  
 % \mbox{\includegraphics[width=0.9\hsize]{kr_fig_x2_persistent_switch_p.eps}}  
\caption{Poincar\'e section corresponding to the time series shown in  
figure~\ref{fig:x2perswitch}. The inset shows an enlargement of the region in  
the box marked in the main picture. The Poincar\'e section is $H_1^\smallin$  
with $|x_3|=h=0.01$. A dot (resp.~cross) indicates that the orbit next crosses  
the Poincar\'e section with $x_2>0$ (resp.~$x_2<0$). The inset shows that  
crosses (indicating a switch) occur where the trajectory lies  
below~${\cal W}^\smalls(P)$ (approximately $x_2=0$).}  
 \label{fig:x2perswitchp}  
\end{figure}

\begin{figure}  
\centerline{\epsfxsize=0.9\hsize\epsfbox{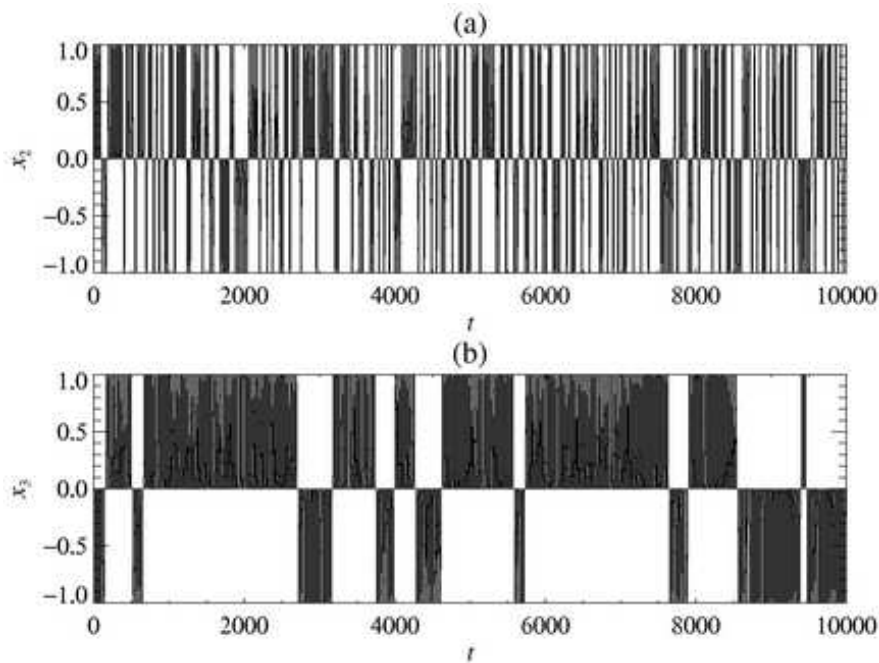}}  
 % \mbox{\includegraphics[width=0.9\hsize]{kr_fig_x2_x3_switch.eps}}  
\caption{Time series showing persistent switching in $x_2$ and $x_3$, for  
(\ref{eq:modelSBODEsfirst}--\ref{eq:modelSBODEslast}) with  $\epsilon_1=0.005$,  
$\epsilon_2=3\times10^{-5}$, $\epsilon_3=0.001$. Other coefficients as defined in  
text.}  
 \label{fig:x2x3switch}  
\end{figure}

\begin{figure}  
\centerline{\epsfxsize=0.9\hsize\epsfbox{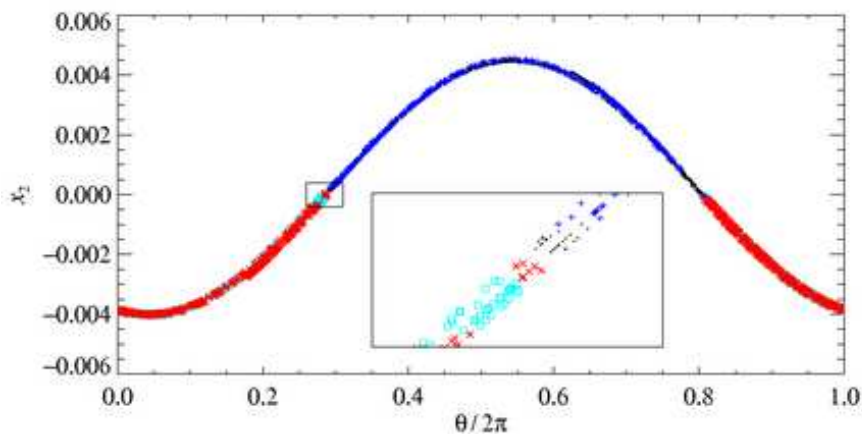}}  
 % \mbox{\includegraphics[width=0.9\hsize]{kr_fig_x2_x3_switch_p.eps}}  
\caption{Poincar\'e section corresponding to the time series shown in  
figure~\ref{fig:x2x3switch}. The inset shows an enlargement of the region in  
the box marked in the main picture. The Poincar\'e section is $H_1^\smallin$  
with $|x_3|=h=0.01$. Four symbols are used: a dot (resp.~cross) indicates that  
the orbit next crosses the Poincar\'e section with $x_2>0$ (resp.~$x_2<0$) and  
with $x_3>0$. A~$+$ (resp.~square) indicates that the orbit next crosses the  
Poincar\'e section with $x_2>0$ (resp.~$x_2<0$) and with $x_3<0$. The division  
between orbits falling either side of ${\cal W}^\smalls(P)$ is clearly visible. Orbits  
falling on opposite sides of ${\cal W}^\smalls(\pm E_2)$ are reasonably well mixed with this  
choice of cross-section.}  
 \label{fig:x2x3switchp}  
\end{figure}

The behaviour for negative values of $\epsilon_2$ and/or $\epsilon_3$ is  
analogous: single switches can be found, but there is no persistent switching  
for $\epsilon_1=10^{-4}$. However, persistent switching in one or both of $x_2$  
and $x_3$ is observed if we increase $\epsilon_1$.  Figures~\ref{fig:x2perswitch} and~\ref{fig:x2perswitchp} show an example of   
persistent switching in~$x_2$ (but not~$x_3$) for $\epsilon_1=3\times10^{-4}$,   
$\epsilon_2=2\times10^{-4}$ and $\epsilon_3=0.001$. The trajectory crosses the   
Poincar\'e section in a curve that appears reasonably smooth at the largest   
scale (figure~\ref{fig:x2perswitchp}), but the magnified inset shows that the  
curve has structure, and that parts of the curve lie below~${\cal W}^\smalls(P)$, leading   
to switches from $x_2>0$ to $x_2<0$. For   
these parameter values, there appears to be no attractor with $x_2<0$,   
and after a short transient, the trajectory switches back to $x_2>0$.   
Dynamics with a larger value of $\epsilon_1=0.005$ is shown in  
figures~\ref{fig:x2x3switch} and~\ref{fig:x2x3switchp}: here we have persistent  
switching in $x_2$ and~$x_3$. 
  
 %%%%%%%%%%%%%%%%  
 %%%%%%%%%%%%%%%%%  
  
 %%%%%%%%%%%%%%%%  
 %%%%%%%%%%%%%%%%%  
   
 %%%%%%%%%%%%%%%%  
 %%%%%%%%%%%%%%%%%  
  
 %%%%%%%%%%%%%%%%  
 %%%%%%%%%%%%%%%%%  

 %%%%%%%%%%%%%%%%  
\begin{figure}  
\centerline{\epsfxsize=0.9\hsize\epsfbox{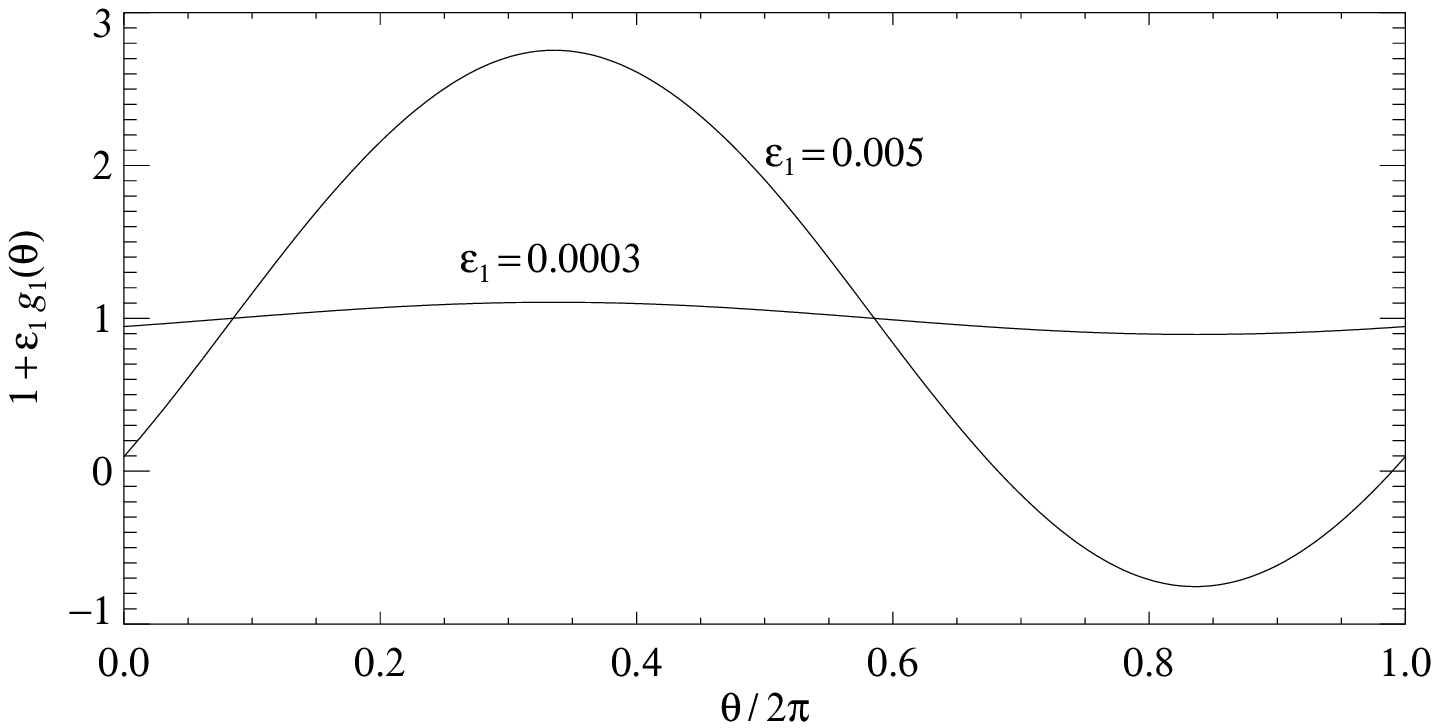}}  
\caption{The function $1+\epsilon_1g_1(\theta)$ fitted to the  
data in figures~\ref{fig:x2perswitch} ($\epsilon_1=3\times10^{-4}$)  
and~\ref{fig:x2x3switch} ($\epsilon_1=0.005$). This function must be both   
positive and negative as a function of~$\theta$ in order to allow persistent   
switching in~$x_3$.}  
 \label{fig:x3fittedmap}  
\end{figure}  
 %%%%%%%%%%%%%%%%%  
  
Our understanding of the mechanism behind persistent switching in $x_2$ or  
$x_3$ requires that $\epsilon_1$ be large enough that $1+\epsilon_1g_3(\theta)$  
or $1+\epsilon_1g_1(\theta)$ can take on positive and negative values, as a  
function of~$\theta$. In order to illustrate this effect, we have fitted these  
functions using the trajectories in figures~\ref{fig:x2perswitch}  
($\epsilon_1=3\times10^{-4}$) and~\ref{fig:x2x3switch} ($\epsilon_1=0.005$). We have  
concentrated on switching in~$x_3$, so the fact that these trajectories are for  
different values of~$\epsilon_2$ does not affect our conclusions. We took the   
coordinates $(x_3,\theta_1)$ on the section~$H_1^\smallout$ and $\tilde{x}_3$   
from the next intersection with~$H_2^\smallin$, and used these data to fit a   
map of the form of~$\Psi_{12}$:  
 $$  
 \tilde{x}_3= x_3 \left(A_1 + \epsilon_1 f_1(\theta_1)\right)  
              + \epsilon_3 \left(1 + \epsilon_1g_1(\theta_1)\right),  
 $$  
see equation~(\ref{eq:psi12}). We represented the two functions $f_1$ and~$g_1$  
by a finite Fourier series, and were able to fit the data to within one part in  
1000 for all points with the smaller value of~$\epsilon_1$, and to within one  
part in 100 for all but a handful of the points for the larger value  
of~$\epsilon_1$. As expected, the $A_1+\epsilon_1f_1$ part of the map remains  
positive, but $1+\epsilon_1g_1$ can change sign for the larger value  
of~$\epsilon_1$, as shown in figure~\ref{fig:x3fittedmap}. Indeed, the  
numerical ratio of the amplitudes of the two fitted functions is $16.663$ while  
the ratio of the two values of $\epsilon_1$ is $16.667$. Our understanding  
requires the change in sign of $1+\epsilon_1g_1$ as a necessary condition for  
persistent switching in~$x_3$, which is confirmed by this illustration and by  
our other calculations. A similar transition occurs (at a smaller value  
of~$\epsilon_1$) at the onset of persistent switching in~$x_2$, and in that  
case, the data can be fitted to within 1 part in 10,000 or better.  
  
We note that we were able to find parameter values associated with persistent  
switching in~$x_2$ alone but were unable to get persistent switching in $x_3$  
without also having switching in $x_2$. This is a consequence of the particular  
choice of symmetry-breaking coefficients we use: the threshold in $\epsilon_1$  
for persistent switching in $x_3$ is higher than the threshold for persistent  
switching in $x_2$ for the chosen coefficients (since we have $d_{23}>d_{33}$).  
For other parameter choices, the   
thresholds could be the other way around.  
  
In the numerical simulations described above, the quantities $\delta_1$,  
$\delta_2$, $\delta_3$ and $\delta$ were all greater than one. This choice was  
made to ensure that that the heteroclinic cycle was attracting in the fully  
symmetric case and to remove any possible complications due to chaotic dynamics  
associated with homoclinic bifurcations of $\pm E_2$ and $\pm E_3$. Much of the  
same switching dynamics will still occur if one or both of $\delta_2$ and  
$\delta_3$ is less than one while $\delta>1$, since the mechanism for switching 
we have found does not depend on the size of the individual $\delta_i$. However, in  
this case there may be  
additional complications in the dynamics associated with the homoclinic  
bifurcations of the equilibria.  
  
\section{Conclusions}  
\label{sec:conclusions}  
  
This paper has investigated the effect of small symmetry-breaking on the  
dynamics near a structurally stable heteroclinic cycle connecting two  
equilibria and a periodic orbit. The heteroclinic cycle is structurally stable  
in the case that there are two reflection symmetries and a rotation symmetry in  
the underlying system; we were interested in the dynamics seen when one or more  
of the symmetries is broken. It was reported in~\cite{MPR} that this type of  
system can exhibit seemingly chaotic dynamics along with repeated but irregular  
switching of sign of various variables, but details of the mechanisms  
underlying the onset of complicated dynamics were not explored there. In this  
paper, we have identified global bifurcations that induce the onset of chaotic  
dynamics and switching near a heteroclinic cycle of this type. These turn out  
to be homoclinic tangencies between the stable and unstable manifolds of the  
periodic orbit, and specific heteroclinic tangencies between stable and  
unstable manifolds of the two equilibria. By construction and analysis of  
approximate return maps, we were able to locate (approximately) the global  
bifurcations in parameter space and hence to isolate instances of the different  
types of switching and chaotic dynamics in a specific numerical example.  
  
In addition to identification of the mechanisms underlying the onset of  
switching, two important insights have been gained from this study. First, we  
found that interaction of the different symmetry-breaking terms is required for  
switching; partial symmetry breaking (where one or two of the three  symmetries  
are retained) did not result in switching. Switching results from the right  
combination of a global bifurcation (which results in turn from breaking of the  
rotation symmetry) and small breaking of at least one of the reflection  
symmetries. Second, we found there is a threshold in $\epsilon_1$ below which  
there can be single switches in the signs of certain variables but no  
persistent switching. The important point here is that persistent switching  
does not result from arbitrarily small symmetry breaking, but is a `large'  
symmetry-breaking effect. Of course, `small' and `large' are relative  
terms, and addition of seemingly tiny symmetry-breaking effects might actually  
result in persistent switching, as was the case in the numerical example we  
investigated in section~\ref{sec:numerics}.  
  
One aspect of this problem which has not yet been investigated is whether it is  
possible to make {\it a priori} predictions about switching rates or derive  
scaling laws for switching times. It is plausible that switching rates and  
times might depend on the `distance' from the global bifurcation that induces  
the switching, but no detailed attempts have yet been made to quantify such a  
relationship. The statistics of switching intervals were measured in the  
related model of~\cite{MPR}, who report an exponential distribution of  
intervals between switches.  
  
Finally, we note that the dynamo model in~\cite{MPR} has a symmetry 
that is never broken (this is the symmetry $(z_1,x_2,x_3) \to (-z_1,-x_2,-x_3)$ 
in the notation of~\cite{MPR}). Retention of this symmetry while breaking all 
others amounts to retaining invariance of the $z_1=x_2=x_3=0$ subspace, and 
will have a consequence of relating the dynamics in different parts of 
the phase space. For example, if it is possible to switch from $(x_2>0,x_3>0)$ 
to $(x_2>0,x_3<0)$, it will also be possible to switch from $(x_2<0,x_3<0)$ to 
$(x_2<0,x_3>0)$. Our results do not include this effect, and retaining this 
symmetry may well have profound effects on the switching properties. 
Nevertheless we expect our basic ideas about switching being induced by a 
balance between a global bifurcation and symmetry-breaking terms and the 
existence of a threshold for persistent switching to apply quite generally, and 
to the example in~\cite{MPR} in particular, even if the details turn out not to  
be directly relevant. 
 
\section{Appendix: Details of return map construction}   
  
\subsection{Coordinates and cross-sections}  
 \label{sec:coords}  
  
Following~\cite{KrMe95}, we distinguish radial, contracting, and expanding  
directions near the equilibria in the fully symmetric case.  
If ${\mathcal P}_1=\{(z_1, x_2, x_3): x_3=0\}$, ${\mathcal P}_2=\{(z_1,  
x_2, x_3): z_1=0\}$, ${\mathcal P}_3=\{(z_1, x_2, x_3): x_2=0\}$, with  
${\mathcal P}_0\equiv {\mathcal P}_{3}$, then the radial eigenvalues at $\pm  
E_j$ ($j=2, 3$) are the eigenvalues of the linearised vector field at $\pm E_j$  
(i.e., eigenvalues of $(d{\mathbf f})_{\pm E_j}$) restricted to ${\mathcal P}_j  
\cap {\mathcal P}_{j-1}$. The contracting eigenvalues are the remaining  
eigenvalues of $(d{\mathbf f})_{\pm E_j}$ in ${\mathcal P}_{j-1}$, and the  
expanding eigenvalues are the remaining eigenvalues in ${\mathcal P}_{j}$. The  
radial direction is then the span of the eigenvectors corresponding to the  
radial eigenvalues, and similarly for the contracting and expanding directions.  
Near $P$ we define the radial direction to be the direction of ${\mathcal P}_1  
\cap {\mathcal P}_{3}$ (i.e., the plane $x_2=x_3=0$), the contracting direction  
is parallel to the $x_3$-axis, and the expanding direction is parallel to the  
$x_2$-axis. These definitions are consistent with those in~\cite{KrMe95}   
but are adapted for the case where there is a  
periodic orbit in the heteroclinic cycle.  
  
We choose local coordinates near each of $P$, $\pm E_2$, and~$\pm E_3$ to 
make the linearised dynamics as simple as possible. Near $+E_2$ in the fully  
symmetric case, we define $\xi_2=x_2-\bar{x}_2$, where $\bar{x}_2$ is the value  
of $x_2$ at~$+E_2$, and then use local coordinates $(z_1,\xi_2,x_3)$; $z_1$,  
$\xi_2$ and $x_3$ correspond to the contracting, radial and expanding  
directions, respectively. Under symmetry breaking, $+E_2$ moves in proportion to  
the magnitude of the symmetry breaking, and the local coordinates are measured  
from the new position of the equilibrium point. The eigenvalues and  
eigenvectors change similarly, but since the eigenvalues are generically  
distinct and non-zero, small symmetry-breaking will not change the nature of  
the local structure and we can use the slightly altered eigenvectors to define  
a slightly altered local coordinate system. We continue to identify radial,  
contracting and expanding directions once weak symmetry breaking is introduced,  
in the obvious way, and retain the notation $(z_1,\xi_2,x_3)$, for the altered  
coordinates, although $z_1$ and $x_3$ may no longer coincide with the  
corresponding global coordinates.  
  
A similar construction is used near $-E_2$ except that $\xi_2=-x_2+\bar{x}_2$,  
where $\bar{x}_2$ is the value of $x_2$ at~$-E_2$. The point of defining  
$\xi_2$ in this way is that positive values of $\xi_2$ near $+E_2$ are mapped  
under the reflection $\kappa_2$ to positive values of $\xi_2$ near $-E_2$, and  
this simplifies the maps we derive below. An analogous procedure is used to  
define local coordinates near $\pm E_3$.  
  
To construct local coordinates near $P$, we select a cross-section transverse  
to $P$, say $\theta_1=0$. Near~$P$, the flow induces a map from that section to  
itself, with $P$ corresponding to a fixed point of the map. We define  
$\xi_1=r_1-\bar{r}_1$, where $\bar{r}_1$ is the value of $r_1$ at the fixed  
point; $\xi_1$ is the analogue in the map to the radial coordinate for the flow  
near $P$. The remaining local coordinates  on the cross-section are defined by  
restricting the expanding and contracting directions at $P$, as defined above,  
to the cross-section. Local coordinates can be extended to a neighbourhood of  
the whole of~$P$  in the  
fully symmetric case by applying equivariance under $\kappa_1$. Finally, small  
symmetry-breaking perturbations will not change the local structure near $P$,  
and we can extend to slightly altered local coordinates $(\xi_1, \theta_1, x_2,  
x_3)$ in a neighbourhood of $P$ so long as we remember that symmetry-breaking  
terms may have a different effect at each value of $\theta_1$, so for instance,  
$z_1=\bar{r}_1{\rm e}^{{\rm i}\theta_1}$ where $\bar{r}_1 \equiv r_1(\theta_1  
)$ is a function of $\theta_1$. Note that the global polar coordinates $(r_1,  
\theta_1)$ are well-defined near $P$ even in the presence of small symmetry  
breaking since $P$ is far from the origin.  
   
Cross-sections in~$\Rset^4$ are defined in terms of local coordinates as  
follows:  
 \begin{align*}  
 H_1^\smallin  &= \left\{(\xi_1,\theta_1,x_2,x_3): |\xi_1| \leq h,  
                                           0 \leq \theta_1 < 2\pi,  
                                           |x_2|\leq h,  
                                           |x_3|=h \right\},\\  
 H_1^\smallout &= \left\{(\xi_1,\theta_1,x_2,x_3): |\xi_1| \leq h,  
                                           0 \leq \theta_1 < 2\pi, |x_2|=h,  
                                           |x_3|\leq h \right\},\\  
 H_2^\smallin  &= \left\{(z_1,\xi_2,x_3): |z_1|=h,  
                                           |\xi_2|\leq h,  
                                           |x_3|\leq h \right\},\\  
 H_2^\smallout &= \left\{(z_1,\xi_2,x_3): |z_1|\leq h,  
                                           |\xi_2| \leq h,  
                                           |x_3|=h \right\},\\  
 H_3^\smallin  &= \left\{(z_1,x_2,\xi_3): |z_1|\leq h,  
                                           |x_2|=h,  
                                           |\xi_3|\leq h \right\},\\  
 H_3^\smallout &= \left\{(z_1,x_2,\xi_3): |z_1|=h,  
                                           |x_2|\leq h,  
                                           |\xi_3|\leq h\right\}.  
 \end{align*}  
The cross-sections $H_2^{\smallin}$ and $H_2^{\smallout}$  
(resp.~$H_3^{\smallin}$ and $H_3^{\smallout}$) work equally well near $\pm E_2$  
(resp.~$\pm E_3$) so long as the local coordinate $\xi_2$ (resp.~$\xi_3$) is  
interpreted correctly, as described above.  
  
We also define a Poincar\'e section for the periodic orbit~$P$:  
 \begin{equation*}  
 H_1^P = \left\{(\xi_1,\theta_1,x_2,x_3): |\xi_1| \leq h,  
                                           \theta_1=0,  
                                           |x_2|\leq h,  
                                           |x_3|\leq h \right\}.  
 \end{equation*}  
Trajectories visiting~$P$ first cross~$H_1^\smallin$, may then cross $H_1^P$  
several times, and eventually leave the neighbourhood of~$P$ on  
crossing~$H_1^\smallout$.  
  
\subsection{Local maps}  
  
Within a neighbourhood of each of $\pm E_2$, $\pm E_3$ and $P$, so long as  
certain non-resonance conditions on the eigenvalues are satisfied, the dynamics  
can be linearised using the Hartman--Grobman theorem~\cite{GH86}. In the fully  
symmetric case, the dynamics near $P$ can be approximated by:  
 $$\dot\xi_1= -2\xi,\qquad  
   \dot\theta_1= 1, \qquad  
   \dot x_2= e_1 x_2,\qquad  
   \dot x_3= -c_1 x_3,$$  
 where $e_1$ and $c_1$ are positive constants. Without loss of  
generality, we have assumed that the radial eigenvalue  is~$-2$, and that the angular speed is 
1. Solving these equations, we find the local map $\phi_1:H_1^\smallin\to H_1^\smallout$  
is given by:  
 \begin{equation}  
 \phi_1(\xi_1,\theta_1,x_2,x_3)=\left(  
            \xi_1\left|\frac{x_2}{h}\right|^{\gamma_1},  
            \theta_1 -\frac{1}{e_1} \ln \left|\frac{x_2}{h}\right|,  
            h\sgn(x_2),  
            h\sgn(x_3)\left|\frac{x_2}{h}\right|^{\delta_1}\right)\,,  
 \label{eq:phi1map}  
 \end{equation}  
where the initial value of $x_3$ satisfies $|x_3|=h$, where $\sgn(x)=+1$ if $x>0$,  
$\sgn(x)=-1$ if $x<0$, and $\sgn(0)=0$, and where  
$\delta_1=c_1/e_1$, $\gamma_1=2/e_1$.  
  
The argument that symmetry-breaking does not affect this local map  
goes as follows. The transition from $H_1^\smallin\to H_1^\smallout$  
has three parts. First, the trajectory travels from  
$H_1^\smallin\to H_1^P$ in less than one circuit around~$P$. The trajectory  
does not get very close to~$P$ in this time, having started at least a  
distance~$h$ from it. Since the $\epsilon_i$'s, which control the symmetry  
breaking, are assumed to be much smaller than~$h$, the fully symmetric flow  
yields an adequate approximation of the true flow. Second, the trajectory makes  
$n_1$ circuits around the periodic orbit from $H_1^P$ to $H_1^P$, where $n_1$  
is a non-negative integer no greater than $T_1/2\pi$. These circuits are  
governed by the linearised Poincar\'e map and its Floquet multipliers: ${\rm  
e}^{-4\pi}$, ${\rm e}^{2\pi e_1}$ and ${\rm e}^{-2\pi c_1}$ in the radial,  
expanding and contracting directions, respectively, where, to leading order in  
the~$\epsilon_i$'s, the period of~$P$ is~$2\pi$. The number~$n_1$  
is unchanged by the weakly broken symmetry, and so, to leading order, this part  
of the map is unchanged. Third, the trajectory travels from $H_1^P\to  
H_1^\smallout$ in less than one circuit around~$P$ and again is not too close  
to~$P$, so the fully symmetric flow yields an adequate approximate of the true  
flow. Composing these three parts yields~(\ref{eq:phi1map}), to  
leading order.  
  
Local maps $\phi_2:H_2^\smallin\to H_2^\smallout$ and  
$\phi_3:H_3^\smallin\to H_3^\smallout$ are obtained similarly:  
 \begin{equation}  
 \phi_2(r_1=h,\theta_1,\xi_2,x_3)=\left(  
            h\left|\frac{x_3}{h}\right|^{\delta_2},  
            \theta_1  -\frac{1}{e_2}\ln\left|\frac{x_3}{h}\right|,  
            \xi_2\left|\frac{x_3}{h}\right|^{\gamma_2},  
            h\sgn(x_3)\right)\,,  
 \label{eq:phi2map}  
 \end{equation}  
 \begin{equation}  
 \phi_3(r_1,\theta_1,x_2,\xi_3)=\left(  
            h,  
            \theta_1 -\frac{1}{e_3}\ln\left(\frac{r_1}{h}\right) ,  
            h\sgn(x_2)\left(\frac{r_1}{h}\right)^{\delta_3},  
            \xi_3\left(\frac{r_1}{h}\right)^{\gamma_3}\right)\,,  
 \label{eq:phi3map}  
 \end{equation}  
where $c_i$ and $e_i$ are the absolute  
values of the real part of the contracting and expanding eigenvalues  
at~$+E_i$, $\delta_i=c_i/e_i$, $\gamma_i=2/e_i$, and  $|x_2|=h$. As for $\phi_{1}$, the radial eigenvalues 
and the angular speeds are chosen to be $-2$ and $1$.  
 
\subsection{Global maps}  
  
The global map $\Psi_{12}:H_1^\smallout\to H_2^\smallin$ takes orbits from a  
neighbourhood of $P$ to a neighbourhood of $+E_2$. We write  
 \begin{equation*}  
 \Psi_{12}(\xi_1,\theta_1,x_2=h,x_3)=  
 (\tilde{r}_1=h, \tilde{\theta}_1,  
 \tilde{\xi}_2, \tilde{x}_3)  
 \end{equation*}  
and initially do not include symmetry-breaking effects. The unstable  
manifold of $P$ is two-dimensional and, locally, intersects  $H_1^\smallout$ at  
 \begin{equation}  
 {\cal W}^\smallu(P) \cap H_1^\smallout = \left\{(\xi_1,\theta_1,x_2,x_3):  
                                            \xi_1=0,  
                                            0\leq \theta_1< 2\pi,  
                                            x_2=\pm h,  
                                            x_3=0 \right\}\,  
 \label{eq:WuP}  
 \end{equation}  
The manifold ${\cal W}^\smallu(P)$ has two branches: the {\em positive} branch  
intersects $H_1^\smallout$ with $x_2=h$ and the {\em negative} branch  
intersects $H_1^\smallout$ with $x_2=-h$. The positive branch forms a  
connection from $P$ to $+E_2$ and is the solution we now linearise about,  
while the negative branch forms a connection from $P$ to $-E_2$ and will be  
discussed later. The positive branch of ${\cal W}^\smallu(P)$ intersects  
$H_2^\smallin$ at   
 \begin{equation}  
 \left\{(r_1, \theta_1,\xi_2,x_3): r_1=h,  
                  0 \leq \theta_1 < 2\pi,  
                       \xi_2=\bar{\xi}_2,  
                       x_3=0 \right\}\,  
 \label{eq:WuP2}  
 \end{equation}  
where $\bar{\xi}_2$ is a small constant. The $\kappa_1$  
symmetry forces the heteroclinic orbit corresponding to the choice  
$\theta_1$ in (\ref{eq:WuP}) to have an angular component in $H_2^\smallin$ of  
$\theta_1+\bar{\theta}_1$ for some constant $\bar{\theta}_1$, i.e., the global  
map acts on the angle as a rigid rotation.  Furthermore, trajectories that are  
near but not on the unstable manifold of ${\cal W}^\smallu(P)$ have  
$\tilde{\xi}_2$ and $\tilde{x}_3$ depending on the initial $\xi_1$ and $x_3$  
but not on~$\theta_1$, while $\tilde{\theta}_1=\theta_1+\bar{\theta}_1$ where  
$\bar{\theta}_1$ is a function of the initial $\xi_1$ and $x_3$. Equivariance  
under $\kappa_3$ ensures that the subspace $x_3=0$ is invariant, that  
$\tilde{x}_3$ is an odd function of $x_3$, and that ${\bar\theta}_1$ and  
$\tilde{\xi}_2$ are even functions of~$x_3$. (The $\kappa_2$ symmetry has no  
role in determining the form of $\Psi_{12}$ although it can be used to  
construct a map from $P$ to $-E_2$ once $\Psi_{12}$ is known.) Writing a Taylor  
series in the small quantities $\xi_1$ and $x_3$ therefore yields  
 \begin{align*}  
 {\tilde\theta}_1(\xi_1,\theta_1,x_3)=  
 \theta_1 + {\bar\theta}_1({\xi}_1,x_3) &=  
                                \theta_1+{\bar\theta}_1(0,0)+\mbox{h.o.t.},\\  
 \tilde{\xi}_2({\xi}_1,x_3) &=  
                                       \tilde{\xi}_2(0,0)+\mbox{h.o.t.},\\  
 \tilde{x}_3({\xi}_1,x_3) &=  
 \frac{\partial\tilde{x}_3}{\partial x_3}(0,0)\,x_3  
 +\mbox{h.o.t.},  
 \end{align*}  
where \hbox{h.o.t.} denotes higher order terms. Effectively, so long as  
${\bar\theta}_1$ and $\tilde{\xi}_2$ are non-zero, they can be replaced by  
constants, while $\tilde{x}_3$ depends linearly on~$x_3$.   We write  
$A_1=\frac{\partial\tilde{x}_3}{\partial x_3}(0,0)$ and  
$B_1=\tilde{\xi}_2(0,0)$, and note that $A_1>0$ since the region of phase space  
with $x_3>0$ is dynamically invariant.  
  
The effect of weak symmetry breaking on these expressions is as follows.  
First, the symmetry $x_3\to-x_3$ is broken by including  
terms that are odd in $x_3$ in the expressions for $\tilde{\theta}_1$ and  
$\tilde{\xi}_2$, and terms that are even in $x_3$ in the expression for  
$\tilde{x}_3$. We multiply all such terms by an overall  
factor of~$\epsilon_3$, which is a real constant that controls the magnitude of  
the breaking of the $\kappa_3$ symmetry. Then the lowest order contribution to  
$\tilde{\theta}_1$ and $\tilde{\xi}_2$ will be a term in $\epsilon_3x_3$ while  
$\tilde{x}_3$ will pick up a term linear in~$\epsilon_3$. At leading order all  
quadratic terms can be dropped, so the only new term is one linear  
in~$\epsilon_3$ in the expression for~$\tilde{x}_3$. Second, breaking  
the~$\kappa_1$~symmetry will result in a weak dependence of all the  
coefficients on~$\theta_1$, with the dependence being periodic in that  
variable. We introduce the parameter $\epsilon_1$, which is a real constant  
that multiplies all terms that break the $\kappa_1$ symmetry and that controls  
the magnitude of the symmetry-breaking terms. For example, $A_1$ will become  
$A_1+\epsilon_1f_1(\theta)$, with the caveat that this term must remain  
positive, for all $\theta$ and $\epsilon_1$. Third, weakly breaking the  
symmetry $x_2 \to -x_2$ will not affect the form of this map.  
  
Putting all this together results in a map  
 $\Psi_{12}:H_1^\smallout\to H_2^\smallin$:  
 \begin{align}  
 \Psi_{12}(\xi_1,\theta_1,x_2=h,x_3)=\big(&  
            \tilde{r}_1=h,  
            \tilde{\theta}_1=\theta_1 + \Phi_1,  
            \tilde{\xi}_2=B_1,\nonumber\\  
           &\tilde{x}_3=A_1 x_3 + \epsilon_3  
                       + \epsilon_1x_3f_1(\theta_1)  
                       + \epsilon_1\epsilon_3g_1(\theta_1)  
            \big),  
 \label{eq:psi12}  
 \end{align}  
where $\Phi_1$, $A_1$, $B_1$ are constants, and $f_1$, $g_1$ are  
$2\pi$-periodic functions of $\theta_1$. The $\theta_1$ dependence  
cannot be treated using Taylor series expansions, as $\theta_1$ is not a small  
quantity. We explain below why some quadratic terms ($\epsilon_1x_3$ and  
$\epsilon_1\epsilon_3$) need to be kept. 
  
Similarly, a map from $P$ to $-E_2$ can be constructed. This  
has precisely the form of (\ref{eq:psi12}), except that it starts from  
$x_2=-h$. Breaking of the $\kappa_2$~symmetry means coefficients in the  
map will be slightly different but the map is unchanged at leading order.  
  
The map $\Psi_{23}:H_2^\smallout\to H_3^\smallin$ is calculated in a similar  
way. In the fully symmetric case, we linearise about  
$ {\cal W}^\smallu(+E_2)$, which intersects $H_2^\smallout$ at $(z_1=0,\xi_2=0,x_3=h)$ and  
$H_3^\smallin$ at $(z_1=0,\xi_2=h,\xi_3=\bar{\xi}_3)$ where $\bar{\xi}_3$ is a  
small constant. For orbits near $ {\cal W}^\smallu(+E_2)$, the value of  
$\xi_2$ at $H_2^\smallout$ does not influence the final position to leading  
order and $z_1$ at $ H_3^\smallin$ depends linearly on the values of $z_1$ at  
$H_2^\smallout$: $\tilde{z}_1=A_2{\rm e}^{{\rm i}\Phi_2}z_1$ for real constants  
$A_2>0$, $\Phi_2$. If the $\kappa_1$~symmetry is broken,  
$ {\cal W}^\smallu(+E_2)$ leaves $H_2^\smallout$ with $z_1=0$ and arrives at $H_3^\smallin$ with  
$z_1=\tilde\epsilon_1$, where $\tilde\epsilon_1= \epsilon_1(a_\smallr+{\rm  
i}a_\smalli)$ for $a_\smallr$ and $a_\smalli$ real constants and $\epsilon_1$  
as defined earlier. Writing the resulting map in terms of the real and  
imaginary parts of $\tilde{z}_1$:  
 \begin{align}  
 \Psi_{23}(r_1,\theta_1,\xi_2,x_3=h)=\big(&  
            \tilde{x}_1=\epsilon_1a_\smallr + A_2r_1\cos(\theta_1+\Phi_2),  
            \nonumber\\  
           &\tilde{y}_1=\epsilon_1a_\smalli + A_2r_1\sin(\theta_1+\Phi_2),  
            \nonumber\\  
           &\tilde{x}_2=h,  
            \tilde{\xi}_3=B_2\big),  
 \label{eq:psi23}  
 \end{align}  
where $a_\smallr$, $a_\smalli$, $A_2$, $B_2$ and $\Phi_2$ are real  
constants determined by the global flow, and $A_2>0$. As in~$\Psi_{12}$, there  
are $2\pi$-periodic functions of~$\theta_1$ in the map, but here the functions  
are known explicitly because the $z_1$ variable is small throughout the  
transition from $+E_2$ to~$+E_3$, and the dynamics of $z_1$ is well-approximated  
by a scaled rotation. Similar maps can be obtained for the three connections  
$-E_2 \to +E_3$ and $\pm E_2 \to -E_3$; although the coefficients will be  
slightly different in each case, to lowest order we obtain the same map for  
each of the other connections so long as the signs of the $x_3$  
(resp.~$\tilde{x}_2$) components are chosen appropriately on the incoming  
(resp.~outgoing) cross-sections (for example, the map from $-E_2$ to $-E_3$  
will have $x_3=-h$ and $\tilde{x}_2=-h$).  
  
The global map $\Psi_{31}:H_3^\smallout\to H_1^\smallin$ is calculated in a  
similar way:  
 \begin{align}  
 \Psi_{31}(r_1=h,\theta_1,x_2,\xi_3)=\big(&  
           \tilde{\xi}_1=B_3,  
           \tilde{\theta}_1 = \theta_1+\Phi_3,  
           \nonumber\\  
          &\tilde{x}_2 = A_3 x_2 + \epsilon_2  
                       + \epsilon_1x_2f_3(\theta_1)  
                       + \epsilon_1\epsilon_2g_3(\theta_1),  
           \nonumber\\  
          &\tilde{x}_3 = h\big),  
 \label{eq:psi31}  
 \end{align}  
where $A_3$, $B_3$ and $\Phi_3$ are real constants, $f_3$ and $g_3$ are  
$2\pi$-periodic functions of $\theta_1$, and $\epsilon_1$ controls the size of  
the terms that break the $\kappa_1$ symmetry. The parameter $\epsilon_2$  
introduced in $(\ref{eq:psi31})$ is analogous to $\epsilon_3$, and is a real  
quantity that controls the size of all terms that break the $\kappa_2$  
symmetry. Similarly to the case for $A_1$ argued above, we take  
$A_3+\epsilon_1f_3(\theta_1)$ to be positive for all values of $\epsilon_1$. A  
similar map can be defined near the connections from $-E_3$ to $P$, and will,  
to leading order, be identical to (\ref{eq:psi31}) so long as $\tilde{x}_3=h$  
is replaced by $\tilde{x}_3=-h$.  
  
The effect of each of the global maps defined above is, at leading order, to  
rotate the angular variable by an order one amount that is independent of other  
variables, to set the radial variable to a constant, and, in the absence of  
symmetry-breaking, to scale the variable that measures proximity to the cycle.  
Symmetry-breaking enters in two ways. First, it destroys  
the invariant subspaces thus destroying some of the heteroclinic connections  
that made up the cycle. Second, breaking the $\kappa_1$~rotation symmetry  
allows $\theta_1$~dependence to enter into the maps, most importantly through  
the variables $x_3$ in the $\Psi_{12}$~map and $x_2$ in the $\Psi_{31}$~map. It  
is this $\theta_1$ dependence that allows the heteroclinic tangencies discussed  
in Section~\ref{sec:description}.  
 
\subsection{Return maps} 
 
Return maps approximating the dynamics near the heteroclinic cycle can now be 
computed by composing the local and global maps in an appropriate order. For  
instance, to obtain the various forms of the map  
$R :   H_3^\smallin \to  H_3^\smallin$ given by equations  
(\ref{eq:R_nosb}--\ref{eq:R_sb1complex}) we calculate  
$R\equiv \psi_{23} \circ \phi_2 \circ \psi_{12} \circ \phi_1 \circ  \psi_{31} \circ \phi_3$ 
in the usual way.

\section*{Acknowledgments}  
 We thank Ian Melbourne, Edgar Knobloch and Jeff Porter for helpful 
conversations. This research has been supported by grants from New Zealand 
Institute for Mathematics and its Applications, University of Auckland Research 
Council, London Mathematical Society and the Engineering and Physical Sciences 
Research Council.

\end{document}